\documentclass[11pt]{article}
\usepackage{amsfonts,amssymb,amsmath}
\usepackage{stmaryrd}

\def \DATE {December 17, 2007}

\newcommand{\beq}{\begin{equation}}
\newcommand{\eeq}{\end{equation}}
\newcommand{\beqa}{\begin{eqnarray}}
\newcommand{\eeqa}{\end{eqnarray}}
\newcommand{\nn}{\nonumber \\}
\def \podr {&& \hspace{-15pt}}

\def \x {{\mathrm{x}}}
\def \xx {\mbf{\mathrm{x}}}
\def \yy {\mbf{\mathrm{y}}}
\def \zz {\mbf{\mathrm{z}}}
\def \xxi {\xi}

\def \y {{\mathrm{y}}}

\def \rr {{\mathrm{r}}}
\def \qq {{\mathrm{q}}}

\def \ee {{\mathrm{e}}}

\def \Ss {\mathrsfs{S}}

\def \Dd {\mathrsfs{D}}
\def \Ee {\mathrsfs{E}}

\def \R {{\mathbb R}}

\def \Z {{\mathbb Z}}
\def \N {{\mathbb N}}
\def \Sr {{\mathbb S}}

\def \spr {\cdot}
\def \id {\text{\rm id}}
\def \di {\partial}

\def \II {\mathcal{P}}
\def \Jj {\mathcal{R}}
\def \Jp {\widehat{\mathcal{R}}}
\def \JJJ {R}
\def \JPJ {\widehat{R}}
\def \Eps {\Mbf{\varepsilon}}
\def \Lamb {\Mbf{\lambda}}

\def \mz {\bigr\backslash\hspace{1pt}\{0\}}

\def \rdf {\mathcal{Q}}
\def \RDF {Q}

\DeclareMathAlphabet{\mathbbm}{U}{bbm}{m}{n}

\def \Nn {\mathbbm{n}}
\def \Mm {\mathbbm{m}}
\def \Pp {\mathrsfs{T}}

\DeclareSymbolFont{ltrs}     {OT1}{pzc}{m}{it}
\DeclareSymbolFont{ltrsa}     {OMS}{cmsy}{m}{n}
\DeclareSymbolFont{ltrsA}{U}{txmia}{m}{it}
\DeclareSymbolFont{symbolsC}{U}{txsyc}{m}{n}
\DeclareSymbolFont{ltrsB}{U}{rsfs}{m}{n}

\DeclareSymbolFontAlphabet{\mfrak}{ltrsA}
\DeclareMathSymbol{\oo}{\mathord}{ltrsB}{"64}
\DeclareMathAlphabet{\mathpzc}{OT1}{pzc}{m}{it}
\DeclareMathAlphabet{\mathrsfs}{U}{rsfs}{m}{n}
\DeclareMathSymbol{\pord}{\mathrel}{symbolsC}{"1C}
\DeclareMathSymbol{\hdim}{\mathord}{ltrsA}{"64}
\DeclareMathSymbol{\WFU}{\mathord}{ltrsB}{"57}
\DeclareMathSymbol{\MMM}{\mathord}{ltrs}{"4D}
\DeclareMathSymbol{\NNN}{\mathord}{ltrs}{"4E}

\def \ID {\text{\it id}}

\def \Ker {\mathit{Ker}}

\newcommand{\vrestr}[2]{\!\left.\raisebox{#1}{$\,$}\!\right|_{\,
\raisebox{1pt}{\small \(#2\)}}}

\newcommand{\mbf}[1]{\ensuremath{\mathchoice
                    {\mbox{\boldmath$\displaystyle\mathbf{\mathit{#1}}$}}
                    {\mbox{\boldmath$\textstyle\mathbf{\mathit{#1}}$}}
                    {\mbox{\boldmath$\scriptstyle\mathbf{\mathit{#1}}$}}
                    {\mbox{\boldmath$\scriptscriptstyle\mathbf{\mathit{#1}}$}}}}

\newcommand{\Mbf}[1]{\ensuremath{\mathchoice
                    {\mbox{\boldmath$\displaystyle\mathbf{#1}$}}
                    {\mbox{\boldmath$\textstyle\mathbf{#1}$}}
                    {\mbox{\boldmath$\scriptstyle\mathbf{#1}$}}
                    {\mbox{\boldmath$\scriptscriptstyle\mathbf{#1}$}}}}

\newcommand{\Section}[2]{%
 \refstepcounter{section}
 \section*{\large \arabic{section}. #1}\label{#2}%
 \addtocontents{toc}{\protect\vspace{-8pt}}
 \addtocontents{toc}{\contentsline {section}{\thesection.\hspace{6pt}{#1}}{\arabic{page}}}}

\newcounter{Theorem}
\newcounter{Remark}
\newcounter{Definition}
\newcounter{Example}
\newcounter{Exercise}

\def \setcntrs {\setcounter{equation}{0}\setcounter{Theorem}{0}\setcounter{Remark}{0}\setcounter{Definition}{0}\setcounter{Example}{0}\setcounter{Exercise}{0}}

\renewcommand{\thesection}{\arabic{section}}

\newenvironment{Theorem}[1][\bf Theorem \arabic{section}.\arabic{Theorem}]{%
        
        \refstepcounter{Theorem}\noindent\textbf{#1.}${}$\hspace{1pt}${}$\it}{}
\newenvironment{Proposition}[1][\bf Proposition \arabic{section}.\arabic{Theorem}]{%
        
        \refstepcounter{Theorem}\noindent\textbf{#1.}${}$\hspace{1pt}${}$\it}{}
\newenvironment{Lemma}[1][\bf Lemma \arabic{section}.\arabic{Theorem}]{%
        
        \refstepcounter{Theorem}\noindent\textbf{#1.}${}$\hspace{1pt}${}$\it}{}

\newenvironment{Remark}[1][\bf Remark \arabic{section}.\arabic{Remark}]{%
        
        \refstepcounter{Remark}\noindent\textbf{#1.}${}$\hspace{1pt}${}$}{}
\newenvironment{Definition}[1][\bf Definition~\arabic{section}.\arabic{Definition}]{%
        
        \refstepcounter{Definition}\noindent\textbf{#1.}${}$\hspace{1pt}${}$}{}

\newenvironment{Proof}[1][\it Proof]{\noindent\textit{#1.}${}$\hspace{7pt}${}$}{\nolinebreak$\quad$\nolinebreak$\Box$}

\newcounter{tmpc}
\newlength{\tmplenght}
\newlength{\tmplenghta}
\newlength{\tmplenghtb}
\newlength{\tmplenghtc}

\newenvironment{LIST}[1]{%
\setlength{\tmplenghta}{#1}
\setlength{\tmplenghtb}{#1}
\setlength{\tmplenghtc}{#1}
\advance\tmplenghtb-5pt
\advance\tmplenghtc 42pt
\setcounter{tmpc}{0}
\begin{list}{{\rm (\alph{tmpc})}}{\usecounter{tmpc}
\setlength{\leftmargin}{\tmplenghta}
\setlength{\rightmargin}{0cm}
\setlength{\itemsep}{1pt}
\setlength{\topsep}{3pt}
\setlength{\labelsep}{5pt}
\setlength{\labelwidth}{\tmplenghtb}
\setlength{\listparindent}{\tmplenghta}}
}{\end{list}}

\def\DI{\widehat{\Delta}}
\def\Di{\Delta}
\def\DIR{\widetilde{\Delta}}

\def\DP{\Dd'}
\def\CI{\mathcal{C}^{\infty}}
\def\SCI{\mathrsfs{E}}
\def\SCi{\mathrsfs{F}}
\def\du{\hspace{1pt}\bullet}
\def\DSCi{\SCi\raisebox{8pt}{\hspace{0.5pt}}^*}

\def\RM{\mathcal{R}}

\newcommand{\OM}[1]{\mathit{\Omega}_{#1}}
\newcommand{\CLO}[1]{\mathit{\mathcal{Z}}_{#1}}
\newcommand{\EXA}[1]{\mathit{\mathcal{B}}_{#1}}
\def\THETA{\mbf{\mathrm{\Theta}}}
\def\OMEGA{\mbf{\mathrm{\Omega}}}
\def\aLPHA{\mbf{\mathrm{\alpha}}}

\def\BBB{\text{\textbf{\textit{B}}}}

\newcommand{\HOM}[1]{H^{#1}}

\def\wed{^{\hspace{1pt}\wedge}}

\newcommand{\DO}[1]{\mathfrak{D}\raisebox{-3.5pt}{\hspace{-2pt}}_{#1}}

\def\scdeg{\text{\rm sc.{\hspace{1pt}}d.}}

\def \Lccl {\omega}

\def \Sccl {\gamma}
\def \sccl {\Gamma}
\def \bsccl {\Mbf{\Gamma}}

\title{Cohomological analysis of the Epstein-Gla\-s\-er renormalization}

\author{Nikolay M. Nikolov}

\date{\DATE}

\begin{document}

\maketitle

\thispagestyle{empty}

\vspace{-0.8cm}

\begin{center}
\scriptsize
Institute for Nuclear Research and Nuclear Energy, \\
Tsarigradsko Chaussee 72, BG-1784 Sofia, Bulgaria \\
mitov@inrne.bas.bg
\end{center}

\vspace{0cm}

\begin{abstract}
A cohomological analysis of the renormalization freedom
is performed in the Epstein-Glaser scheme on a flat Euclidean space.
We study the deviation from commutativity between the renormalization
and the action of all linear partial differential operators.
It defines a Hochschild $1$--cocycle
and the renormalization ambiguity corresponds
to a \textit{nonlinear} subset in the cohomology class of this renormalization cocycle.
We have shown that the related cohomology spaces
can be reduced to de Rham cohomologies of the so called
``(ordered) configuration spaces''.
We have also found cohomological differential equations
that exactly determine the renormalization cocycles up to the renormalization freedom.
This analysis is a first step towards a new approach for computing renormalization group actions.
It can be also naturally extended to manifolds as well as
to the case of causal perturbation theory.
\end{abstract}

\vspace{0.5cm}

\tableofcontents

\Section{Introduction}{Sect-I}
\setcntrs

Perturbation theory in Quantum Field Theory (QFT) is one of the
technically most difficult subjects in the contemporary theoretical physics.
This is, first and foremost, due to the appearance of complicated integrals in
higher orders, as well as, to the complexity of the accompanying renormalization.
For the realistic QFT models there are practically no numerical results
for arbitrary orders in perturbation theory.
There are also a few methods that allow to perform calculations to all orders.
Without pretending to give justice to various approaches to the subject
we just point out the general analysis of perturbative renormalization
in recent work of Connes-Kreimer (see e.g., \cite{K98,CK00, CK01}).

The present work is a first step to a new approach for determining the action of
the renormalization group in perturbative QFT (i.e., for calculating beta functions).
It offers in addition a geometric insight to the problem.
The general idea of the method is to perform a cohomological analysis
of the renormalization ambiguity and to use it to determine the renormalization group action.
Furthermore, we have separated the problem from the particular models of perturbative QFT,
i.e. we consider all possible theories and even more general situations.
It is this generality that makes the geometric interpretation possible.
It is also important that we do not confine our treatment to the one parameter
action of the renormalization group but consider all linear partial differential operators.
This is done in order to restrict as much as possible the related cohomology owing to the
general properties of the algebra of all differential operators.
Our geometric view favors the study of renormalization in ``coordinate space''.
This approach has been originally developed by Bogolubov, and
Epstein and Glaser \cite{EG73} on Minkowski space and
recently applied to more general pseudo-Riemann manifolds
(see e.g., \cite{BF00, HW}).
It is also called causal perturbation theory.
This approach has a simpler counterpart in Euclidean QFT \cite{GB}.
We choose to work here within this Euclidean framework, and even on
$\R^D$,
but our analysis can be extended to manifolds, as well as to the case of the
causal perturbation theory on pseudo-Riemann manifolds.
Our choice was motivated by the fact that
the geometric structures appearing in the analysis are much more transparent
in the Euclidean approach.

We continue with a more detailed introduction to the subject of the present work.
In Euclidean field theory on $\R^D$ one uses quantities,
like partition functions and correlators, which are formally expressed as
Feynman path integrals.
By the Wick theorem (Gaussian integration)
the path integrals are further reduced in perturbation theory to integrals of
the following general type:
\beq\label{eq1.1}
\mathop{\int}\limits_{\hspace{-10pt}\R^{Dn}} \
\left(\raisebox{14pt}{\hspace{-2pt}}\right.
\mathop{\prod}\limits_{1 \, \leqslant \, j \, < \, k \, \leqslant \, n}
\!\!
G_{jk} (\x_j-\x_k)
\left.\raisebox{14pt}{\hspace{-2pt}}\right) \! \left(\raisebox{14pt}{\hspace{-2pt}}\right.
\mathop{\prod}\limits_{m \, = \, 1}^n
\!
F_m (\x_m)
\left.\raisebox{14pt}{\hspace{-2pt}}\right)
\mathop{\prod}\limits_{\ell \, = \, 1}^n d^D \x_{\ell}\,,
\eeq
where $G_{jk} (\x_j-\x_k)$ are ``propagators''
and $F_m (\x_m)$ are smooth functions on $\R^D$,
which arise from the smearing of the external propagators
($F(\x)$ $=$ $\int G(\x-\y)$ $f(\y)$ $d^D \y$).
The important thing for us is that the propagators
$G_{jk} (\x_j-\x_k)$
are regular functions for $\x_j \neq \x_k$.
So, the integrand in (\ref{eq1.1}) is well defined,
regular function on the subspace of all pairwise distinct arguments
$(\x_1,\dots,\x_n) \in \bigl(\R^D\bigr)^{\times\hspace{1pt}n}( \, \cong$ $\R^{Dn})$.
The latter subspace of $\R^{Dn}$
is also called an \textit{ordered configuration space}
over $\R^D$ and is denoted by $F_n \bigl(\R^D\bigr)$.
The configuration spaces are generally introduced for arbitrary manifold (or set) $X$ by:
\beq\label{FS}
F_n \bigl( X \bigr)
\, = \,
\Bigl\{\bigl(\x_1,\dots,\x_n\bigr) \in X^{\times n} \, : \, \x_j \, \neq \, \x_k \ \ \text{if} \ \ j \, \neq \, k\Bigr\} \,
\eeq
and they are well studied (see e.g., \cite{FH}).

In this way we arrive at the following general problem.
Find an extension of functions belonging to
$\CI \bigl(F_n \bigl(\R^D \bigr)\bigr)$
to distributions over the whole space $\R^{Dn}$.
In fact, since the integrals in (\ref{eq1.1}) are on the whole space $\R^{Dn}$, then
the integrand should be additionally extended
to a linear functional over the vector space of all \textit{bounded} smooth functions.
Such an extension is another problem known as \textit{infrared renormalization},
which we do not consider here
(but it can be treated by the same method).

\medskip

We shall explain now what kind of cohomological analysis we are going to perform.
Let us introduce the rough and total diagonals, $\DI_n ( \, = \DI_n (D))$ and $\Di_n$
in $\bigl(\R^D\bigr){}^{\times n}$ $\cong$ $\R^{Dn}$, respectively:
\beqa\label{eq1.2DIR}
\hspace{-20pt}
\DI_n := \podr
\Bigl\{
(\x_1,\dots,\x_n) \in \R^{Dn}
\, : \,
\x_j = \x_k
\ \, \text{for some} \ \,
1 \leqslant j < k \leqslant n
\Bigr\} \,, \qquad
\\ \label{eq1.2DiR}
\Di_n := \podr
\Bigl\{
(\x_1,\dots,\x_n) \in \R^{Dn}
\, : \,
\x_1 = \cdots = \x_n
\Bigr\} \,
,
\eeqa
and so we have the identity:
\beq\label{eq1.4jjj1}
\R^{Dn} {\hspace{2pt}}\bigl\backslash{\hspace{2pt}} \DI_n
\equiv
F_n \bigl(\R^D \bigr) \,.
\eeq
Consider a linear map
$$
\Jj_n :
\SCI_n
\to \DP \bigl(\R^{Dn}\bigr)\,,
$$
where $\SCI_n$ is some vector subspace of
$\CI \bigl(F_n \bigl(\R^D\bigr)\bigr)$, which is a \textit{differential} $\CI \bigl(\R^{Dn}\bigr)$--module\footnote{%
i.e., closed under taking partial derivatives and multiplication by smooth functions;
later on we shall call it also a $\DO{Dn}$--module according to Definition~\ref{Df3.1zz11}},
and $\DP \bigl(\R^{Dn}\bigr)$ is the distributions space over $\R^{Dn}$.
We require $\Jj_n$ to be a \textit{renormalization map} in the sense that
$$
\Jj_n \bigl(u\bigr) \vrestr{12pt}{F_n \bigl(\R^D\bigr)} \, = \, u \,.
$$
There are additional conditions on $\Jj_n$, which we do not consider at the moment.
If $\Jj_n': \SCI_n \to \DP \bigl(\R^{Dn}\bigr)$
is another map that satisfies the identify
$\Jj_n' \bigl(u\bigr) \vrestr{12pt}{F_n \bigl(\R^D\bigr)}$ $=$ $u$,
then the difference
$$
\rdf_n \, := \, \Jj_n - \Jj_n'
$$
is a map
$\SCI_n$ $\to$ $\DP \bigl[\DI_n\bigr]$,
where $\DP \bigl[\DI_n\bigr]$ is the space of distributions supported at the rough diagonal $\DI_n$.
Conversely, $\Jj_n' := \Jj_n + \rdf_n$ for
$\rdf_n : \SCI_n$ $\to$ $\DP \bigl[\DI_n\bigr]$ satisfies again
$\Jj_n' \bigl(u\bigr) \vrestr{12pt}{F_n \bigl(\R^D\bigr)}$ $=$ $u$.
Thus, the space of maps
$\rdf : \SCI_n$
$\to$ $\DP \bigl[\DI_n\bigr]$
reflects the renormalization ambiguity.

Let us now consider the ``commutator'':
$$
c_n \bigl[A\bigr] := A \circ \Jj_n - \Jj_n \circ A \,:\,
\SCI_n
\to \DP \bigl[\DI_n\bigr] \,,
$$
for any linear partial differential operator $A$ with $\CI \bigl(\R^{Dn}\bigr)$--coefficients.
One easily sees that $c \bigl[A\bigr]$ is a \textit{Hoc\-h\-s\-c\-hild $1$--cocycle}:
$$
A_1 \circ c_n \bigl[A_2\bigr] - c_n \bigl[A_1 \circ A_2\bigr] + c_n \bigl[A_1\bigr] \circ A_2 \, = \, 0 \,
$$
for the algebra of all linear differential operators on $\R^{Dn}$.
On the other hand, considering $c_n' \bigl[A\bigr] := A \circ \Jj_n' - \Jj_n' \circ A$
we obtain that the difference $c_n \bigl[A\bigr]-c_n' \bigl[A\bigr]$ is a
\textit{Hoc\-h\-s\-c\-hild coboundary}:
$$
c_n \bigl[A\bigr]-c_n' \bigl[A\bigr] \, = \, A \circ \rdf_n - \rdf_n \circ A \,.
$$
Hence, the cohomological class of $c_n \bigl[A\bigr]$, does not depend on the renormalization ambiguity.
It is important to stress at this point that the class of $c_n \bigl[A\bigr]$, which
exactly corresponds to the renormalization freedom is actually
a \textit{nonlinear} subset of the Hochschild cohomology class of $c_n \bigl[A\bigr]$.
This is because there is an additional condition on the renormalization maps, which is nonlinear,
and which makes their construction inductive in $n$ (see Sect.~\ref{se3ne1}).

From our analysis (Theorem~\ref{Th3.1ne1}), it follows that the renormalization ambiguity allows us to achieve
$c_n \bigl[f\bigr] = 0$ for smooth functions $f$ (i.e., differential operators of zeroth order).
This allows us to extend our methods also for manifolds without even any metric structure on them.
Moreover, the remaining nontrivial part of the cocycle $c_n \bigl[\di_{x_k^{\mu}}\bigr]$
($\di_{x_k^{\mu}}$ $=$ $\frac{\di}{\di x_k^{\mu}}$, see below for our notations)
can be characterized by certain cohomological equations (Eqs.~(\ref{eq4.6ne2}) and (\ref{eq6.9ne1})).
We have proven in Theorem~\ref{Th4.1qq1} that the ambiguity in the solutions of these equations
exactly corresponds to the renormalization freedom.
This is done by using de Rham cohomologies of configuration spaces.
So, in the subsequent two sections we consider the general problem
of constructing renormalization maps and the possibility
to make them commuting with the multiplication by smooth functions.
Then, we analyze the remaining nontrivial cohomological properties of the renormalization maps
and their reduction to de Rham cohomologies.

\textit{Some conventions and notations.}
$\N = \{1,2,\dots\}$, $\N_0 = \{0\} \cup \N$, $\Z = \{0,\pm1,\pm 2,\dots\}$.
We denote vectors of $\R^D$ (our Euclidean space-time)
by $\x$, $\y$, $\dots$,
with $\x$ $=$ $\bigl(x^{\mu}\bigr){}_{\mu \, = \, 1}^D$ $=$ $\bigl(x^1,\dots,x^D\bigr)$,
and the $n$-tuples of vectors of $\R^{Dn}$ by $\xx$ $=$ $\bigl(x^{\mu}_k\bigr)$, $\yy$, $\dots$.
Sometimes, we also consider
instead of $\R^{Dn}$
the spaces $\R^{D(n-1)}$ or $\R^N$ for arbitrary $N \in \N$
and then we continue to denote their elements by $\xx$ (etc.),
but in the case of $\R^N$ the coordinates are denoted by
$\bigl(x^{\xxi}\bigr){}_{\xxi \, = \, 1}^N ( \, = \xx)$.
The partial derivatives as
$\frac{\di}{\di x^{\mu}}$, $\frac{\di}{\di x^{\xxi}}$, $\frac{\di}{\di x^{\mu}_k}$, $\dots$,
are for short denoted by
$\di_{x^{\mu}}$, $\di_{x^{\xxi}}$, $\di_{x^{\mu}_k}$, $\dots$.
We use multiindex notations like
$\xx^{\rr} := \prod_{\xxi} \bigl(x^{\xxi}\bigr)^{r_{\xxi}}$ and
$\di^{\rr} := \prod_{\xxi} \di_{x^{\xxi}}^{r_{\xxi}}$ for
$\rr=(r_1,\dots,r_N) \in \N_0^N$;\
$|\rr| := \sum_{\xxi} r_{\xxi}$, $\rr! := \prod_{\xxi} r_{\xxi}!$.

The distributions space over an open set $U \subseteq \R^N$
is denoted by $\DP \bigl(U\bigr)$ and the test functions space, by $\Dd \bigl(U\bigr)$.
We write $u \bigl[f(\xx)\bigr]$ for a distribution $u(\xx)$
smeared by a test function $f(\xx)$, i.e. what corresponds to the formal expression
$\int u(\xx) f(\xx) d^N\xx$.
We denote the distributions supported at zero in $\R^N$ by $\DP \bigl[0\bigr]$
and to avoid confusion sometimes we indicate the dimension $N$
by writing $0 \in \R^N$.

\Section{Scaling analysis and primary renormalization maps}{NS2.1}
\setcntrs

We consider the problem of extending distributions from
$F_n \bigl(\R^{D}\bigr)$ to the whole space $\R^{Dn}$
following the analog of the Epstein-Glaser scheme on Euclidean spaces.
In this approach an important role plays the notion
of scaling degree of distributions introduced by Steinmann~\cite{S71}.
It is defined by:
\beq\label{eq3.1xx1}
\scdeg \, u \, := \,
\mathop{\min} \, \bigl\{\lambda \in \Z
\, : \,
\mathop{\text{w-\!}\lim}\limits_{\varepsilon \downarrow 0} \, \varepsilon^{\lambda} \, u (\lambda \xx) \, = \, 0 \bigr\} \, ,
\eeq
for a distribution $u$ $\in$ $\DP \bigl(U\bigr)$ over an open cone $U$ in $\R^N$.
(In general, one can define a real scaling degree but to simplify our statements
we decided to work with the integral one.)
Let us denote
\beqa
\label{eq2.5ne111}
\DP_{(L)} \bigl(U\bigr)
\, := \podr
\bigl\{
u (\xx) \in \DP_t \bigl(U\bigr)
\, : \,
\scdeg \, u \, < \, L,\
u \text{\ has a finite order}
\bigr\} \,, \qquad
\\
\label{eq2.4ne111}
\DP_t \bigl(U\bigr)
\, := \podr
\mathop{\bigcup}\limits_{L \, = \, 0}^{\infty} \,
\DP_{(L)} \bigl(U\bigr) \,,
\eeqa
so that we obtain an increasing sequence of vector spaces,
i.e. an \textit{increasing filtration}.

Note that Eq.~(\ref{eq3.1xx1}) means that if $\lambda \geqslant \scdeg \, u$
then the set
$$
\Bigl\{u \bigl[f \bigl(\varepsilon^{-1} \xx \bigr)\bigr] \, \varepsilon^{-N+\lambda} : \varepsilon \, \in \, (0,1)\Bigr\}
$$
is bounded for every test function $f \in \Dd \bigl(U\bigr)$.
By the Banach-Steinhaus theorem~(\cite{Rud})
it follows that this is equivalent to the existence of a test functions norm $\|$$\cdot$$\|_K$ and a constant $C_K > 0$
for every compact $K \subseteq U$, such that for every $f \in \Dd (U)$ with $\text{\it supp} \, f \subseteq K$
and $\varepsilon$ $\in$ $(0,1)$ we have:
$$
\Bigl|u \bigl[f \bigl(\varepsilon^{-1} \xx \bigr)\bigr] \Bigr| \, \leqslant \,
C_K \, \bigl\|f\bigr\|_K \, \varepsilon^{N-\lambda} \,.
$$

We introduce a similar notion of scaling degree for distributions depending on additional variables.
Let $U$ be an open cone in $\R^N$ and $V$ be an open set $\R^M$.
We say that a distribution $u \in \DP \bigl(U \times V \bigr)$
has \textbf{bounded scaling degree} with respect to the first (vector) argument,
which we shall denote by $\xx$,
iff there exit $\lambda \in \Z$, $C_K > 0$ and a test functions norm $\|$$\cdot$$\|_K$
for every compact $K \subseteq U \times V$, such that
for every $f \in \Dd \bigl(U \times V\bigr)$ with $\text{\it supp} \, f \subseteq K$ and $\varepsilon \in (0,1)$ we have
\beq\label{eq2.1ne31}
\Bigl| u \bigl[f \bigl(\varepsilon^{-1}\xx,\yy\bigr)\bigr] \Bigr|
\, \leqslant \,
C_K \, \bigl\|f\bigr\|_K \, \varepsilon^{N-\lambda} \,.
\eeq
The minimal $\lambda \in \Z$ for which this estimate is valid ia called a \textbf{scaling degree}
with respect to $\xx$ and we denote it by
$$
\scdeg_{(\xx)} \, u.
$$

Note that if $u \in \DP \bigl(\R^N \times V\bigr)$ then
\beq\label{eq2.2ne31}
\scdeg_{(\xx)} \Bigl( u \vrestr{12pt}{U \times V} \Bigr) \, \leqslant \, \scdeg_{(\xx)} \, u \,.
\eeq
Also, for $u \in \DP \bigl(U \times V\bigr)$ and $f \in \CI \bigl(U \times V\bigr)$ we have
\beq\label{eq2.3ne22}
\scdeg_{(\xx)} \, \bigl(f \, u\bigr) \, \leqslant \, \scdeg_{(\xx)} \, f + \scdeg_{(\xx)} \, u \,.
\eeq
Let us denote, as above,
\beqa
\label{eq2.5ne1}
\DP_{(\xx \hspace{1pt}:\hspace{1pt} L)} \bigl(U \times V\bigr)
\, := \podr
\bigl\{
u (\xx,\yy) \in \DP_t \bigl(U \times V\bigr)
\, : \,
\scdeg_{(\xx)} \, u \, < \, L,
\nn
\podr \hspace{7pt}
u \text{\ has a finite order}
\bigr\} \,, \qquad
\\
\label{eq2.4ne1}
\DP_{(\xx)\hspace{1pt}t} \bigl(U \times V\bigr)
\, := \podr
\mathop{\bigcup}\limits_{L \, = \, 0}^{\infty} \,
\DP_{(\xx \hspace{1pt}:\hspace{1pt} L)} \bigl(U \times V\bigr) \,.
\eeqa

The following theorem is the basic step towards renormalization of distributions
and can be called \textit{primary renormalization}.

\medskip

\begin{Theorem}\label{Th2.1ne1}
Let $V$ be an open set in $\R^M$.
There exists a linear map
\beq\label{eq2.7ne1}
\II_N :
\DP_{(\xx)\hspace{1pt}t} \Bigl(\bigl(\R^N \mz\bigr) \times V\Bigr)
\to
\DP \bigl(\R^N \times V\bigr) \,,
\eeq
such that
for every \
\(u(\xx,\yy) \in
\DP_{(\xx)\hspace{1pt}t} \Bigl(\bigl(\R^N \mz\bigr) \times V\Bigr)\)
and every $f$ $\in$ $\CI \bigl(\R^N$ $\times V\bigr)$ we have:%
\beqa\label{eq2.7ne31}
&
\II_N \bigl(u\bigr) \vrestr{12pt}{\bigl(\R^N \mz\bigr) \times V} \, = \, u \,,
& \\ \label{eq2.8ne31} &
\scdeg_{(\xx)} \, \II_N \bigl(u\bigr) \, \leqslant \, \scdeg \, u \,,
& \\ \label{eq2.9ne31} &
\II_N \bigl(f u\bigr) \, = \, f \, \II_N \bigl(u\bigr)\,.
&
\eeqa
In addition, for every diffeomorphism $g: V \to V$ and $\eta = 1,\dots,N$ we have:
\beq
\II_N \bigl(g_* \hspace{1pt} u\bigr)
\, = \,
g_* \hspace{1pt} \II_N \bigl(u\bigr)
\label{eq2.11ne31}
, \quad
\II_N \bigl(\di_{y^{\eta}} \hspace{1pt} u\bigr)
\, = \,
\di_{y^{\eta}} \hspace{1pt} \II_N \bigl(u\bigr) \,,
\eeq
where $(g_* \hspace{1pt} u) \bigl[f (\xx,\yy)\bigr]$ $:=$ $u \bigl[f \bigl(\xx,g (\yy)\bigr)\bigr]$.
\end{Theorem}

\medskip

We start the \textit{proof} of this theorem by a lemma
that corresponds to an old result due to Steinmann.

\medskip

\begin{Lemma}\label{Lm2.2ne1}
Let $\Ee$ be the subspace of $\DP_{(\xx)\hspace{1pt}t} \Bigl(\bigl(\R^N \mz\bigr) \times V\Bigr)$,
which consists of those elements $u$ such that $\scdeg_{(\xx)} u < N$
$($i.e., $\Ee$ $=$ $\DP_{(\xx \hspace{1pt}:\hspace{1pt} N-1)} \Bigl(\bigl(\R^N \mz\bigr)$ $\times V\Bigr)$$)$.
Then there exists a unique linear map
\beq\label{eq2.7ne2}
\II_{N,0} : \Ee
\to
\DP_t \bigl(\R^N \times V\bigr) \,,
\eeq
which satisfies the
conditions~(\ref{eq2.7ne31})--(\ref{eq2.11ne31})
of Theorem~\ref{Th2.1ne1}
(for $\II_{N,0}$ instead of $\II_N$ and $u \in \Ee$).
\end{Lemma}

\medskip

The proof of this lemma can be done following the arguments in \cite[Thoerem 2]{BF99}.
For the sake of completeness we just give the definition of $\II_{N,0}$.
Introduce a test function $\vartheta (\xx) \in \Dd \bigl(\R^N\bigr)$
that is equal to $1$ in a neighbourhood of $0$ and set
\beq\label{eq2.8ne21}
\II_{N,0} \bigl(u\bigr) \, := \,
\mathop{\text{w-\!}\lim}\limits_{n \to \infty} \,
u (\xx,\yy) \bigl(1 - \vartheta \bigl(2^n \xx\bigr) \bigr) \,.
\eeq
The limit exists in $\DP \bigl(\R^N\bigr)$ since for every test function
$f \in \DP \bigl(\R^N\bigr)$ the sequence
$$
\bigl(u (\xx,\yy) \bigl(1 - \vartheta \bigl(2^n \xx\bigr)\bigr)\bigr) \bigl[ f(\xx,\yy) \bigr] \equiv
\int \! u (\xx,\yy) \bigl(1 - \vartheta \bigl(2^n \xx\bigr)\bigr) \, f(\xx,\yy) \, d^N \! x\, d^M \! y
$$
is fundamental.

Continuing with the proof of Theorem~\ref{Th2.1ne1}
we define
linear
maps
\beq\label{eq2.9ne21}
\II_{N,L} : \DP_{(\xx \hspace{1pt}:\hspace{1pt} N+L-1)} \Bigl(\bigl(\R^N \mz\bigr) \times V\Bigr) \to
\DP_{(\xx \hspace{1pt}:\hspace{1pt} N+L-1)} \bigl(\R^N \times V\bigr)
\eeq
for $L \geqslant 0$,
using $\II_{N,0}$ in the following way.
Taking again a test function $\vartheta (\xx) \in \Dd \bigl(\R^N\bigr)$
that is equal to $1$ in a neighbourhood of $0$ we introduce for
test functions $f (\xx,\yy) \in \Dd \bigl(\R^N \times V\bigr)$
the \textit{truncated} Taylor remainder
\beq\label{eq2.10ne21}
\mathop{\sum}\limits_{|\rr|=L} \xx^{\rr} \, T_{\rr} \bigl(f\bigr) (\xx,\yy)
\, = \,
f (\xx,\yy)
\, - \,
\mathop{\sum}\limits_{|\qq| < L} \, \frac{1}{\qq!} \, \bigl(\di^{\qq} f\bigr) (0,\yy) \, \xx^{\qq} \, \vartheta (\xx) \,,
\eeq
so that $T_{\rr} (f) \in \Dd \bigl(\R^N \times V\bigr)$.
Then we set
\beq\label{eq2.11ne21}
\II_{N,L} \bigl(u\bigr) \bigl[f\bigr] \, = \,
\mathop{\sum}\limits_{|\rr|=L}
\II_{N,0} \bigl(\xx^{\rr} \hspace{1pt} u\bigr) \bigl[T_{\rr} \bigl(f\bigr) (\xx,\yy)\bigr] \,.
\eeq
The so defined $\II_{N,L}$ are renormalization maps in the sense that
they satisfy the condition (\ref{eq2.7ne31}).
But still, these maps are not consistent since
\beq\label{eq2.18ne32}
\II_{N,L+1}\vrestr{12pt}{\DP_{(\xx \hspace{1pt}:\hspace{1pt} N+L-1)}
\bigl(\bigl(\R^N \mz\bigr)\times V\bigr)} \, \neq \,
\II_{N,L} \,.
\eeq
In fact,
\beq\label{eq2.18ne31}
\II_{N,L+1} \bigl(u\bigr) - \II_{N,L} \bigl(u\bigr) =
\mathop{\sum}\limits_{|\rr|=L} \frac{(-1)^{|\rr|+1}}{\rr!} \
\II_{N,0} \bigl(\xx^{\rr} \hspace{1pt} u (\xx,\yy)\bigr) \bigl[ \vartheta (\xx)\bigr]
\, \delta^{(\rr)} (\xx) \,,
\eeq
for $u \in \DP_{(\xx \hspace{1pt}:\hspace{1pt} N+L-1)} \Bigl(\bigl(\R^N \mz\bigr)\times V\Bigr)$,
where $\delta^{(\rr)}$ $:=$ $\di^{\rr} \delta$ and
$\II_{N,0} \bigl(\xx^{\rr} \hspace{1pt} u\bigr) \bigl[ \vartheta (\xx)\bigr]$
is understood as an element of $\DP \bigl(V\bigr)$.
Consider the linear maps
\beq\label{eq2.19ne31}
\alpha_{\rr,0} :
u \, \mapsto \, \II_{N,0} \bigl(\xx^{\rr} \hspace{1pt} u\bigr) \bigl[ \vartheta (\xx)\bigr] :
\DP_{(\xx \hspace{1pt}:\hspace{1pt} N+L-1)} \Bigl(\bigl(\R^N \mz\bigr)\times V\Bigr)
\to
\DP \bigl(V\bigr)\,
\eeq
for $\rr \in \N_0^N$ such that $|\rr|=L$.
There always exist extensions
\beq\label{eq2.20ne31}
\alpha_{\rr} : \DP_{(\xx) \hspace{1pt} t} \Bigl(\bigl(\R^N \mz\bigr)\times V\Bigr)
\to
\DP \bigl(V\bigr)
\eeq
of these maps, i.e.,
\(\alpha_{\rr}
\vrestr{12pt}{\DP_{(\xx \hspace{1pt}:\hspace{1pt} N+L-1)} \bigl(\bigl(\R^N \mz\bigr)\times V\bigr)}\)
$=$ $\alpha_{\rr,0}$ and even more:

\medskip

\begin{Lemma}\label{Lm-add}
The extensions~(\ref{eq2.20ne31}) can be chosen to satisfy also the properties
\beq\label{eq2.21.ne31}
\alpha_{\rr} \circ g^* \, = \, g^* \circ \alpha_{\rr} \,,
\quad
\alpha_{\rr} \circ \di_{y^{\eta}} \, = \, \di_{y^{\eta}} \circ \alpha_{\rr} \,,
\eeq
for every diffeomorphism $g: V \to V$ and $\eta = 1,\dots,N$.
\end{Lemma}

\medskip

We omit the proof of this lemma. The main idea is to equip the vector space
$\DP_{(\xx) \hspace{1pt} t} \Bigl(\bigl(\R^N \mz\bigr)\times V\Bigr)$ with a suitable topology
such that
\beq\label{eq2.24ne44}
\DP_{(\xx) \hspace{1pt} t} \Bigl(\bigl(\R^N \mz\bigr)\times V\Bigr)
\ \cong \
\DP_{t} \bigl(\R^N \mz\bigr)
\ \widehat{\otimes} \
\DP \bigl(V\bigr)
\eeq
($\widehat{\otimes}$ being a topological tensor product)
and then construct $\alpha_{\rr}$ to act only on the first term in~(\ref{eq2.24ne44}).
We point out that apart from this lemma we do not use
any continuity assumptions about our linear maps.

\medskip

Hence, the maps
\beq\label{missed1}
\II_{N,L}' \, := \,
\II_{N,L} -
\mathop{\sum}\limits_{|\rr| \, \leqslant \, L} \, \frac{(-1)^{|\rr|+1}}{\rr!} \ \delta^{(\rr)} (\xx) \,
\cdot
\alpha_{\rr}
\vrestr{12pt}{\DP_{(\xx \hspace{1pt}:\hspace{1pt} N+L-1)} \bigl(\bigl(\R^N \mz\bigr)\times V\bigr)}
\eeq
are consistent, i.e., they satisfy Eq.~(\ref{eq2.18ne32}).
Thus, if we set
\beq\label{qqq}
\II_N'
\vrestr{12pt}{\DP_{(\xx \hspace{1pt}:\hspace{1pt} N+L-1)} \bigl(\bigl(\R^N \mz\bigr)\times V\bigr)}
\, := \, \II_{N,L}'
\eeq
we obtain a linear map $\II_N'$ that yields all
the properties~(\ref{eq2.7ne31}), (\ref{eq2.8ne31}),
and (\ref{eq2.11ne31}).
It remains only to establish the existence of a map $\II_N$ that satisfies also (\ref{eq2.9ne31}).

To achieve this we apply the construction in the following lemma.

\medskip

\begin{Lemma}\label{Lm2.4ne1}
There exists a linear map~(\ref{eq2.7ne1})
that satisfies the properties~(\ref{eq2.7ne31}), (\ref{eq2.8ne31})
and (\ref{eq2.11ne31})
of Theorem~(\ref{Th2.1ne1}), and the equalities
\beq\label{eq2.9ne1}
\II_N \bigl(x^{\xi} \hspace{1pt} u\bigr) \, = \, x^{\xi} \, \II_N \bigl(u\bigr)
\eeq
for every $\xi=1,\dots,N$, where $\xx = \bigl(x^1,\dots,x^{N}\bigr)$.
\end{Lemma}

\medskip

\begin{Proof}
By the above considerations there exists a linear map $\II_N'$~(\ref{eq2.7ne1})
satisfying the properties (\ref{eq2.7ne31}), (\ref{eq2.8ne31}),
and (\ref{eq2.11ne31}).
If $\II_N : \DP_{(\xx)\hspace{1pt}t} \bigl(\bigl(\R^N \mz\bigr) \times V\bigr)$
$\to$ $\DP_{(\xx)\hspace{1pt}t} \bigl(\R^N \times V\bigr)$ is another map that satisfies
(\ref{eq2.7ne31}), (\ref{eq2.8ne31}),
and (\ref{eq2.11ne31})
we set
\beqa\label{eq2.15ne21}
\rdf := \II_N' - \II_N : \podr
\DP_{(\xx)\hspace{1pt}t} \bigl(\bigl(\R^N \mz\bigr) \times V\bigr)
\to
\DP \bigl[0\bigr] \otimes \DP \bigl(V\bigr) \,,
\\ \label{eq2.16ne21}
\hspace{-30pt}
c_{\xi} := x^{\xi} \circ \II_N' - \II_N' \circ x^{\xi} : \podr
\DP_{(\xx)\hspace{1pt}t} \bigl(\bigl(\R^N \mz\bigr) \times V\bigr)
\to
\DP \bigl[0\bigr] \otimes \DP \bigl(V\bigr)
\eeqa
($\xi = 1,\dots,N$),
where $\DP \bigl[0\bigr]$ is the vector space of distributions
supported at $0 \in \R^N$.
Then Eq.~(\ref{eq2.9ne1}) is equivalent to
\beq\label{eq2.17ne21}
c_{\xi} \, = \, x^{\xi} \circ \rdf - \rdf \circ x^{\xi} \,.
\eeq
Thus, the problem is to find a linear map $\rdf$~(\ref{eq2.15ne21}) such that
it does not raise the scaling degree in $\xx$ and (\ref{eq2.17ne21}) is satisfied.

To this end we expand $\rdf$ and $c_{\xi}$ in delta functions and their derivatives:
\beqa\label{eq3.11}
\hspace{-30pt}
\rdf = \!\podr \mathop{\sum}\limits_{\rr \, \in \, \N_0^N} \ \frac{1}{\rr!} \
\delta^{(\rr)} (\xx)
\, \RDF_{\rr}
\, , \quad
\RDF_{\rr} : \DP_{(\xx)\hspace{1pt}t} \bigl(\bigl(\R^N \mz\bigr) \times V\bigr)
\to \DP \bigl(V\bigr),
\\ \label{eq2.19ne21}
\hspace{-30pt}
c_{\xi} = \!\podr \mathop{\sum}\limits_{\rr \, \in \, \N_0^N} \ \frac{1}{\rr!} \
\delta^{(\rr)} (\xx)
\, C_{\xi,\rr}
\, , \quad
C_{\xi,\rr} : \DP_{(\xx)\hspace{1pt}t} \bigl(\bigl(\R^N \mz\bigr) \times V\bigr)
\to \DP \bigl(V\bigr).
\eeqa
The condition that $\rdf$ does not raise the scaling degree is equivalent to
\beq\label{eq3.12}
\RDF_{\rr} \, u \, = \, 0
\quad \text{if} \quad
|\rr| \, > \, \scdeg \, u - N \,,
\eeq
and Eq.~(\ref{eq2.17ne21}) holds iff
\beq\label{eq3.13}
\RDF_{\rr+\ee_{\xxi}} \bigl[u\bigr] \, = \,
- \, C_{\xxi,\rr} \bigl[u\bigr]
-
\RDF_{\rr} \bigl[x^{\xxi} \, u\bigr] \,
\eeq
for all $\rr \in \N_0^N$,
where
$\ee_{\xxi}$ is the $\xxi$th basic vector in $\R^N$
(this follows from the representations (\ref{eq2.19ne21}) and (\ref{eq3.11}), and
the formula $x^{\xxi} \, \delta^{(\rr)} (\xx)$ $=$ $-r_{\xxi}$ $\delta^{(\rr - \ee_{\xxi})} (\xx)$).

The maps $c_{\xi}$ satisfy an ``integrability'' relation
\beq\label{eq3.14n}
c_{\xi} \circ x^{\eta} - x^{\xi} \circ c_{\eta} \, = \,
c_{\eta} \circ x^{\xi} - x^{\eta} \circ c_{\xi} \,,
\eeq
which implies
\beq\label{eq3.15n}
C_{\xi,\rr+\ee_{\eta}} \bigl[u\bigr] - C_{\xi,\rr} \bigl[x^{\eta} \hspace{1pt} u\bigr]
\, = \,
C_{\eta,\rr+\ee_\xi} \bigl[u\bigr] - C_{\eta,\rr} \bigl[x^{\xi} \hspace{1pt} u\bigr] \,.
\eeq
The fact that $\II_N'$ does not raise the scaling degree implies that
$\scdeg \, \bigl(c_{\xi} \, u\bigr) \leqslant - 1 + \scdeg \, u$
(because of Eq.~(\ref{eq2.3ne22})) and then we obtain
\beq\label{eq2.24ne21}
C_{\xi,\rr} \, u \, = \, 0 \,
\quad \text{if} \quad
|\rr| > \scdeg \, u - 1 - N \,.
\eeq

Then let us set
\beq\label{eq3.14}
\RDF_{\rr} \, u
\, := \,
\mathop{\sum}\limits_{\xi \, = \, 1}^N \,
\mathop{\sum}\limits_{s \, = \, 1}^{r_{\xi}} \,
(-1)^{|\qq(\xi,s)|} \,
C_{\xi,\rr-\qq(\xi,s)} \bigl[\xx^{\qq(\xi,s)-\ee_{\xi}} \hspace{1pt} u\bigr] \,,
\eeq
where
$\qq(\xi,s) := s \, \ee_{\xi} + \mathop{\textstyle \sum}\limits_{\eta \, = \, 1}^{\xi-1} r_{\eta} \, \ee_{\eta}$
(writing a sum $\Bigl(\mathop{\sum}\limits_{j \, = \, a}^b \cdots \Bigr)$ with $a,b \in \Z$
we set it zero if $a>b$).
Note that the so defined $\RDF_{\rr}$ satisfy the condition (\ref{eq3.12})
since Eq.~(\ref{eq2.24ne21}) implies that
$C_{\xi,\rr-\qq(\xi,s)} \bigl[\xx^{\qq(\xi,s)-\ee_{\xxi}} \hspace{1pt} u\bigr] = 0$%
~if
$$
|\rr-\qq(\xi,s)| \, > \, \scdeg \, \bigl( \xx^{\qq(\xi,s)-\ee_{\xi}} \, u \bigr) - 1 - N
\quad \Longleftarrow \quad
|\rr| \, > \, \scdeg \, u - N\,
$$
(in the last step we have used Eq.~(\ref{eq2.3ne22})).
Equation~(\ref{eq3.13}) is also satisfied, because of (\ref{eq3.15n}).
Thus, $\RDF_{\rr}$~(\ref{eq3.14}) determine a map $\rdf$ such that Eq. (\ref{eq2.17ne21}) holds
and then
$\II_N := \II_N' - \rdf$ fulfills the conditions of the theorem.
\end{Proof}

\medskip

Now, to complete the proof of Theorem~\ref{Th2.1ne1}
it remains only to show that the above constructed linear map
satisfies Eq.~(\ref{eq2.9ne31}).
By Eq.~(\ref{eq2.9ne1}) it follows that
\beq\label{eq2.38ne31}
\II_N \bigl(\xx^{\rr} \hspace{1pt} u\bigr) \, = \, \xx^{\rr} \, \II_N \bigl(u\bigr)
\eeq
for every $\rr \in \N_0^n$.
Writing then
$$
f(\xx,\yy) \, = \,
\mathop{\sum}\limits_{|\rr|=L} \xx^{\rr} \,g_{\rr}(\xx,\yy)
\, + \,
\mathop{\sum}\limits_{|\qq| < L} \, \frac{1}{\qq!} \, \bigl(\di^{\qq} f\bigr) (0,\yy) \, \xx^{\qq} \,,
$$
we have
$$
\II_N \bigl(f u\bigr) \, = \,
\mathop{\sum}\limits_{|\rr|=L}
\II_{N,0} \Bigl(g_{\rr} (\xx,\yy) . \bigl(\xx^{\rr} \, u\bigr)\Bigr)
+
\mathop{\sum}\limits_{|\qq| < L} \, \frac{1}{\qq!} \, \bigl(\di^{\qq} f\bigr) (0,\yy) \, \xx^{\qq}
\II_N \bigl(u\bigr) \,,
$$
but
$$
\II_{N,0} \Bigl(g_{\rr} (\xx,\yy) . \bigl(\xx^{\rr} \, u\bigr)\Bigr)
\, = \,
g_{\rr} (\xx,\yy) \,
\II_{N,0} \bigl(\xx^{\rr} \, u\bigr)
\, = \, \xx^{\rr} \, g_{\rr} (\xx,\yy) \, \II_N \bigl(u\bigr)
$$
and thus, we obtain (\ref{eq2.9ne31}).

This completes the proof of Theorem~\ref{Th2.1ne1}.

\Section{Inductive renormalization}{se3ne1}
\setcntrs

In this section we consider extensions
of distributions over configuration spaces.
The construction is inductive in the number of points.
In this connection we need a certain multiscale analysis
and we begin with its exposition.

A \textbf{binary tree partition} of $\Nn := \{1,\dots,n\}$ is
a set $\Pp$ of \textit{nonempty} subsets of $\Nn$, such that it contains
$\Nn$, and for every $S \in \Pp$
there are exactly two distinct $S_1,S_2$ $\in$ $\Pp \cup \{1\} \cup \cdots \{n\}$,
which have nonempty intersections with $S$ and then $S$ $=$ $S_1$ $\dot{\cup}$ $S_2$
(in particular, $S$ has at least two elements, i.e. $|S| \geqslant 2$).
It follows that the set $\Pp$ is a binary tree as a partially ordered set
with respect to the inclusion $\subseteq$ and
the elements $S_1,S_2$ $\in$ $\Pp \cup \{1\} \cup \cdots \{n\}$ above are the descents of $S \in \Pp$.
Note also that we always have $|\Pp|=n-1$.

For a binary tree partition $\Pp$ of $\Nn$ and
for every $S \in \Pp$
that has descents $S_1$ and $S_2$ in $\Pp \cup \{1\} \cup \cdots \{n\}$
we set:
\beq\label{eq3.1ne31}
\x_{\Pp,S} \, := \,
\x_{\max S_1} - \x_{\max S_2}
\eeq
(for $\x_1,\dots,\x_n \in \R^D$)
provided that $\max S_1 < \max S_2$
(otherwise,
$\x_{\Pp,S} := \x_{\max S_2} - \x_{\max S_1}$).
There is a linear one-to-one correspondence
$$
\bigl(\x_1,\dots,\x_n\bigr)
\, \longleftrightarrow \,
\bigl(\{\x_{\Pp,S}\}_{S \, \in \, \Pp},\, \x_n\bigr) \,.
$$
To write the inverse passage we denote by $\Pp'(k)$, for $k=1,\dots,n-1$, the subset of $\Pp$
such that $\{k\} \cup \Pp'(k)$
is the \textit{increasing part} of the path connecting
$\{k\}$ and $\{n\}$
in the tree $\Pp \cup \{1\} \cup \cdots \{n\}$.
If $\{k\}$ is a descent of $S$ in $\Pp \cup \{1\} \cup \cdots \{n\}$ we next set
$$
\Pp(k) := \Pp'(k) \quad \text{if} \quad k < \max \, S
\quad \text{and} \quad \Pp(k) := \Pp'(k) \bigl\backslash \{S\}, \quad \text{otherwise} \,.
$$
We then have
\beq\label{eq3.2ne31}
\x_k \, = \, \x_n +
\mathop{\sum}\limits_{S \, \in \, \Pp(k)} \,
\x_{\Pp,S} \,.
\eeq

Let us give an example.
A tree partition of $\Nn$ can be defined via a configuration of brackets
over a permutation $(j_1,\dots,j_n) \in \Ss_n$.
For instance,
$$
\bigl(\bigl(2,3\bigr),\bigl(1,\bigl(4,5\bigr)\bigr)\bigr)
$$
corresponds to the partition
$$
\Pp \, = \, \bigl\{\{1, \dots, 5\}, \{1, 4, 5\}, \{2, 3\}, \{4, 5\}\bigr\}.
$$
For this partition we have
\beqa
&
\x_{\Pp,\{1,\dots,5\}} \, = \, \x_3 - \x_5
,\quad
\x_{\Pp,\{3,4,5\}} \, = \, \x_1 - \x_5
, & \nn &
\x_{\Pp,\{2,3\}} \, = \, \x_2 - \x_3
,\quad
\x_{\Pp,\{4,5\}} \, = \, \x_4 - \x_5
. & \nonumber
\eeqa

For a binary tree partition $\Pp$ of $\Nn$ we set
\beqa\label{eqeq3.3ne31}
\podr
\Eps_{\Pp} \, := \, \bigl(\varepsilon_{\Pp,S}\bigr)_{S \, \in \, \Pp} \, \in \, \R^{\Pp}
,\quad
\Eps_{\Pp}^{N-\Lamb_{\Pp}} \, := \,
\mathop{\prod}\limits_{S \, \in \, \Pp} \, \varepsilon_{\Pp,S}^{N-\lambda_{\Pp,S}} \,,
\nn
\podr
\hspace{-30pt}
\Eps_{\Pp} \cdot \xx \, := \, \bigl(\x_1',\dots,\x_n'\bigr)
\quad \text{for} \quad
\x_k' \, := \,
\x_n + \!
\mathop{\sum}\limits_{S \, \in \, \Pp(k)} \!
\left(\raisebox{12pt}{\hspace{-2pt}}\right.
\mathop{\prod}\limits_{S' \supseteq S} \, \varepsilon_{\Pp,S'}
\left.\raisebox{12pt}{\hspace{-2pt}}\right)
\x_{\Pp,S} \,
\eeqa
($\xx = \bigl(\x_1,\dots,\x_n\bigr) \in \R^{Dn}$).
Then if $\xx \in F_n \bigl(\R^D\bigr)$ and $\varepsilon_{\Pp,S} \in (0,1]$
it follows that $\Eps_{\Pp} \cdot \xx \in F_n \bigl(\R^D\bigr)$.
The geometric meaning of $\Eps_{\Pp} \cdot \xx$
is that it presents a contraction of the configuration of points
$\xx \in F_n \bigl(\R^D\bigr)$ along the tree partition $\Pp$.

Let us consider now a distribution
$u (\xx_1,\dots,\xx_s)$ $\in$
$\DP \Bigl(\mathop{\prod}\limits_{r \, = \, 1}^s F_{n_r} \bigl(\R^{D_r}\bigr)\Bigr)$.
We say that it has a \textbf{tempered growth at the boundary} if
for every binary tree partitions $\Pp_r$ of $\Nn_r$ $(r=1,\dots,s)$
there exit $\Lamb_{\Pp_r,S} \in \Z$ ($S\in \Pp_r$), a constant $C_K$ and
a test functions norm $\|$$\cdot$$\|_K$
for every compact $K \subseteq \mathop{\prod}\limits_r F_{n_r} \bigl(\R^{D_r}\bigr)$, such that
for every $f \in \Dd \Bigl(\mathop{\prod}\limits_r F_{n_r} \bigl(\R^{D_r}\bigr)\Bigr)$,
with $\text{\it supp} \, f \subseteq K$, and $\varepsilon_{\Pp_r,S} \in (0,1)$ we have:
\beq\label{eq3.4ne31}
\Bigl|
u \bigl[f\bigl((\Eps_{\Pp_1})^{-1} \!\cdot \xx_1,\dots,(\Eps_{\Pp_s})^{-1} \!\cdot \xx_s \bigr)\bigr]
\Bigr|
\, \leqslant \,
C_K \, \|f\|_K \, \mathop{\prod}\limits_{r} \Eps_{\Pp_r}^{N-\Lamb_{\Pp_r}} \,,
\eeq
where $(\Eps_{\Pp})^{-1} \!\cdot \xx$ stands for the inverse of the map $\xx \mapsto \Eps_{\Pp} \cdot \xx$.
We denote~by
\beq\label{eq3.5ne31}
\DP_t \Bigl(\mathop{\prod}\limits_r F_{n_r} \bigl(\R^{D_r}\bigr)\Bigr)
\,
\eeq
the subspace of
$\DP \Bigl(\mathop{\prod}\limits_r F_{n_r} \bigl(\R^{D_r}\bigr)\Bigr)$,
which consists of all distributions that have a tempered growth at the boundary
and have finite order.

For every $2$--partition, $\Nn = S \,\dot{\cup}\, S^c$, with nonempty $S$ and $S^c$
we introduce the open set
\beq\label{eq3.7ne2}
V_{\{S,S^c\}} \, = \, \bigl\{\xx=\bigl(\x_1,\dots,\x_n\bigr) \in \R^{Dn}
: \x_j \neq \x_k \text{\ if \ } j \in S,\, k \in S^c\bigr\}
\,.
\eeq
This gives an open covering
\beq\label{eq3.8ne3}
\R^{Dn} \bigr\backslash\hspace{1pt} \Di_n \, = \,
\mathop{\bigcup}\limits_{\mathop{}\limits^{S \, \subset \, \Nn}_{S \, \neq \, \emptyset}} \,
V_{\{S,S^c\}}\,.
\eeq
A straightforward corollary of the above definitions is the following lemma.

\medskip

\begin{Lemma}\label{Lm3.1ne2}
Let $u (\xx,\zz_1,\dots,\zz_s)$ $\in$
$\DP_t \Bigl(F \bigl(\R^D\bigr) \times \mathop{\prod}\limits_r F_{n_r} \bigl(\R^{D_r}\bigr)\Bigr)$
and $\chi (\xx)$ $\in$ $\CI \bigl(\R^{Dn}\bigr)$
be such that $supp \, \chi \subseteq V_{\{\Mm,\Mm^c\}}$
$(\Mm=\{1,\dots,m\},$ $0<m<n)$
then
$$
\chi (\xx) \, u (\xx,\zz_1,\dots,\zz_s)
\, \in \,
\DP_t \Bigl(F_m \bigl(\R^D\bigr) \times F_{n-m} \bigl(\R^D\bigr)
\times \mathop{\prod}\limits_r F_{n_r} \bigl(\R^{D_r}\bigr)\Bigr) \,.
$$%
\end{Lemma}%

\medskip

\begin{Theorem}\label{Th3.1ne1}
There exist linear maps
\beq\label{eq3.1ne1}
\Jj_n :
\DP_t \Bigl(F_n \bigl(\R^D\bigr) \times \mathop{\prod}\limits_r F_{n_r} \bigl(\R^{D_r}\bigr)\Bigr)
\to
\DP_t \Bigl(\R^{Dn} \times \mathop{\prod}\limits_r F_{n_r} \bigl(\R^{D_r}\bigr)\Bigr) \,,
\eeq
for $n=2,3,\dots$,
such that:%
\begin{LIST}{31pt}
\item[$(a)$]
\(
\Jj_n \hspace{1pt}\circ\hspace{1pt} \sigma^* = \sigma^* \hspace{1pt}\circ\hspace{1pt} \Jj_n
\),\hspace{8pt}for every permutation\ $\sigma \in \mathcal{S}_n$,\hspace{4pt}where
$\bigl(\sigma^* u\bigr)\bigl(\xx,$ $\zz_1,$ $\dots,$ $\zz_s\bigr)$ $:=$
$u \bigl(\sigma(\xx),$ $\zz_1,$ $\dots,$ $\zz_s\bigr)$
and
$\sigma(\xx) := \bigl(\x_{\sigma_1},\dots,\x_{\sigma_s}\bigr)$\raisebox{11pt}{}.
\item[$(b)$]
If
$u$ $\in$
$\DP_t \Bigl(F_m \bigl(\R^D\bigr) \times F_{n-m}\bigl(\R^D\bigr) \times \mathop{\prod}\limits_r F_{n_r} \bigl(\R^{D_r}\bigr)\Bigr)$
for $m < n$, then
\beq\label{eq3.2ne1}
\Jj_{m} \bigl(\Jj_{n-m} \bigl(u\bigr)\big) =
\Jj_n \Bigl(u\vrestr{12pt}{F_n \bigl(\R^D\bigr) \times \mathop{\prod}\limits_r F_{n_r} \bigl(\R^{D_r}\bigr)}\Bigr) \,.
\eeq%
\item[$(c)$]
$\scdeg \, \Jj_n \bigl(u\bigr) \, \leqslant \, \scdeg \, u$,\
for every\
$u \in \DP_t \Bigl(F_n \bigl(\R^D\bigr) \times \mathop{\prod}\limits_r F_{n_r} \bigl(\R^{D_r}\bigr)\Bigr)$.
\item[$(d)$]
$\Jj_n \bigl(f u\bigr) \, = \, f \, \Jj_n \bigl(u\bigr)$,\
for every\
$u \in \DP_t \Bigl(F_n \bigl(\R^D\bigr) \times \mathop{\prod}\limits_r F_{n_r} \bigl(\R^{D_r}\bigr)\Bigr)$ and
$f \in \CI \Bigl(\R^{Dn} \times \mathop{\prod}\limits_r F_{n_r} \bigl(\R^{D_r}\bigr)\Bigr)$\raisebox{11pt}{}.
\end{LIST}
\end{Theorem}

\medskip

\begin{Remark}\label{Rm3.1ne1}
$(a)$
The maps $\Jj_n$
depend also on $\mathop{\prod}\limits_r F_{n_r} \bigl(\R^{D_r}\bigr)$
but like in Theorem~\ref{Th2.1ne1} we indicate only the dependence on $n$
for the sake of simplicity of the notation.

$(b)$
The composition in the left hand side of Eq.~(\ref{eq3.2ne1}) is explicitly the following
\beqa\label{eq3.3ne2}
&& \DP_t \Bigl(F_m \bigl(\R^D\bigr) \times F_{n-m}\bigl(\R^D\bigr)
\times \mathop{\prod}\limits_r F_{n_r} \bigl(\R^{D_r}\bigr)\Bigr)
\nn
&& \hspace{10pt} \mathop{\longrightarrow}\limits^{\Jj_{n-m}} \,
\DP_t \Bigl(F_m \bigl(\R^D\bigr) \times \R^{D(n-m)}
\times \mathop{\prod}\limits_r F_{n_r} \bigl(\R^{D_r}\bigr)\Bigr)
\nn
&& \hspace{12pt} \mathop{\longrightarrow}\limits^{\Jj_m} \,\,
\DP_t \Bigl(\R^{Dm} \times \R^{D(n-m)}
\times \mathop{\prod}\limits_r F_{n_r} \bigl(\R^{D_r}\bigr)\Bigr) \,.
\eeqa
Combining this property with $(a)$ and $(c)$ one easily proves by induction in $n$ that
\beq\label{eq3.4ne2}
\Jj_n \bigl(u\bigr)\vrestr{12pt}{F_n \bigl(\R^D\bigr) \times \mathop{\prod}\limits_r F_{n_r} \bigl(\R^{D_r}\bigr)}
\, = \, u \,.
\eeq
\end{Remark}

\medskip

\noindent
\textit{Proof of Theorem~\ref{Th3.1ne1}.}
We use an induction in $n = 2,3,\dots$.
For $n=2$ the theorem reduces to Theorem~\ref{Th2.1ne1} since
\(F_2 \bigl(\R^D\bigr)
= \R^{2D} \bigr\backslash\hspace{1pt} \Di_2
\cong \bigl(\R^D \mz\bigr) \times \R^D\)
and we have to extend distributions from
$\R^D \mz$ to $\R^D$
(this is with respect to the distance $\x_1-\x_2$),
depending also on additional variables whose domain is not extended.

Assume we have proven the theorem for $n' < n$.
We construct then, as a first step, a linear map
\beq\label{eq3.3ne1}
\Jp_n :
\DP_t \Bigl(F_n \bigl(\R^D\bigr) \hspace{0pt} \times \mathop{\prod}\limits_r F_{n_r} \bigl(\R^{D_r}\bigr)\Bigr)
\to
\DP_t \Bigl(\bigl(\R^{Dn} \bigr\backslash\hspace{1pt} \Di_n \bigr) \hspace{0pt}
\times \mathop{\prod}\limits_r F_{n_r} \bigl(\R^{D_r}\bigr)\Bigr) \,,
\eeq
that is uniquely determined by conditions analogous to $(a)$ -- $(d)$:
\begin{LIST}{31pt}
\item[$(a')$]
\(
\Jp_n \hspace{1pt}\circ\hspace{1pt} \sigma^* = \sigma^* \hspace{1pt}\circ\hspace{1pt} \Jp_n
\)\hspace{8pt}($\sigma \in \mathcal{S}_n$);
\item[$(b')$]
If
\(u \in \DP_t \Bigl(F_m \bigl(\R^D\bigr) \times F_{n-m}\bigl(\R^D\bigr)
\times \mathop{\prod}\limits_r F_{n_r} \bigl(\R^{D_r}\bigr)\Bigr)\)
for $m < n$, then
\beqa\label{eq3.6ne1}
\podr
\Jj_{m} \bigl(\Jj_{n-m} \bigl(u\bigr)\big)
\vrestr{12pt}{\bigl(\R^{Dn}\bigr\backslash\Di_n\bigr) \times \mathop{\prod}\limits_r F_{n_r} \bigl(\R^{D_r}\bigr)}
\nn \podr = \,
\Jp_n \Bigl(u\vrestr{12pt}{F_n \bigl(\R^D\bigr) \times \mathop{\prod}\limits_r F_{n_r} \bigl(\R^{D_r}\bigr)}\Bigr)\,;
\eeqa%
\item[$(c')$]
$\scdeg \, \Jp_n \bigl(u\bigr) \, \leqslant \, \scdeg \, u$;
\item[$(d')$]
$\Jp_n \bigl(f u\bigr) \, = \, f \, \Jp_n \bigl(u\bigr)$,\
for every\
\(f \in \CI \Bigl(\bigl(\R^{Dn} \bigr\backslash\hspace{1pt} \Di_n \bigr)
\times \mathop{\prod}\limits_r F_{n_r} \bigl(\R^{D_r}\bigr)\Bigr)\)\raisebox{11pt}{}.
\end{LIST}

Indeed,\raisebox{12pt}{}
let us take a partition of unity $\bigl\{\chi_{\{S,S^c\}}\bigr\}_{S \,\subset\, \Nn}$
that is subordinate to the open covering~(\ref{eq3.8ne3}) and
for every
$S = \{s_1,\dots,s_{m}\}$, with $s_1 < \cdots < s_m$, and
$S^c = \{s_1',\dots,s_{n-m}'\}$, with $s_1' < \cdots < s_{n-m}'$, let us define a permutation
$\sigma_S \in \Ss_n$ by
\beqa\label{s_s}
\hspace{-20pt}
\bigl(1,\dots,n\bigr) \, \mathop{\longmapsto}\limits^{\sigma_S^{-1}} \podr
\bigl(s_1,\dots,s_{m},s'_1,\dots,s'_{n-m}\bigr) \quad \text{if} \quad \max S < \max S^c,
\nn
\hspace{-20pt}
\bigl(1,\dots,n\bigr) \, \mathop{\longmapsto}\limits^{\sigma_S^{-1}} \podr
\bigl(s'_1,\dots,s'_{n-m},s_1,\dots,s_{m}\bigr) \quad \text{if} \quad \max S^c < \max S.
\eeqa
Then we set
\beqa\label{eq3.8ne2}
\hspace{-40pt}
\Jp_n\bigl(u\bigr)
\podr := \!\!
\mathop{\sum}\limits_{\mathop{}\limits^{\emptyset \, \neq \, S \, \subset \, \Nn}_{\max S \, < \, \max S^c}}
\Jj_S \circ \Jj_{S^c}
\bigl(
\chi_{\{S,S^c\}} \hspace{1pt} u
\bigr)
\vrestr{12pt}{\bigl(\R^{Dn}\bigr\backslash\Di_n\bigr) \times \mathop{\prod}\limits_r F_{n_r} \bigl(\R^{D_r}\bigr)}
, \quad
\\ \label{eq3.9ne2}
\Jj_S \podr := \, \sigma_S^* \circ \Jj_m \circ \sigma_S^{*-1} \,.
\eeqa
This defines $\Jp_n (u)$ as an element of
\(\DP_t \Bigl(\bigl(\R^{Dn} \bigr\backslash\hspace{1pt} \Di_n \bigr)
\times \mathop{\prod}\limits_r F_{n_r} \bigl(\R^{D_r}\bigr)\Bigr)\),
because of the induction and Lemma~\ref{Lm3.1ne2}.
The map $\Jp_n$, so defined, automatically satisfies the above conditions $(a')$~--~$(d')$.
The property $(d')$ implies the independence of $\Jp_n$ on the partition of unity
$\bigl\{\chi_{\{S,S^c\}}\bigr\}_{S \,\subset\, \Nn}$ and its uniqueness.

The second, final, step in the construction of $\Jj_n$ is to apply again Theorem~\ref{Th2.1ne1} and take the composition
\beq\label{eq3.7ne1}
\Jj_n \, := \, \II_{D(n-1)} \circ \Jp_n \,,
\eeq
where $\II_{D(n-1)}$ is the linear map:
\beqa\label{eq3.8ne1}
\II_{D(n-1)} : \podr
\DP_t \Bigl( \bigl(\R^{Dn} \bigr\backslash\hspace{1pt}\Di_n\bigr) \times \mathop{\prod}\limits_r F_{n_r} \bigl(\R^{D_r}\bigr)\Bigr)
\nn
\podr \cong  \,
\DP_t \Bigl( \bigl(\R^{D(n-1)} \mz\bigr) \times \R^D \times \mathop{\prod}\limits_r F_{n_r} \bigl(\R^{D_r}\bigr)\Bigr)
\nn
\podr \to \,
\DP_t \Bigl(\R^{Dn} \times \mathop{\prod}\limits_r F_{n_r} \bigl(\R^{D_r}\bigr)\Bigr) \,,
\eeqa
provided by Theorem~\ref{Th2.1ne1}.
The properties $(b)$ -- $(d)$ of Theorem~\ref{Th3.1ne1} are then ensured by the construction.
To obtain the property $(a)$ one should choose $\II_{D(n-1)}$ to commute with all $\sigma^*$,
which can be done by a symmetrization:
$\II_{D(n-1)}$ $=$ $\sum_{\sigma} \sigma^* \circ \II_{D(n-1)}' \circ \sigma^{* -1}$ if we have started with
$\II_{D(n-1)}'$ that does not satisfy the symmetry.$\quad\Box$

\medskip

\begin{Remark}\label{Rm-x1}
We see by the proof of Theorem~\ref{Th3.1ne1} that the properties
of the renormalization maps $\Jj_n$ are completely determined by the properties
of the primary renormalization maps $\II_N$.
The same is true
for the case of the causal perturbation theory on pseudo-Riemann manifolds
with the only difference being the distinct inductive procedure of the renormalization.
It is then more complicated and is performed on (time-ordered, or retarded) products of fields
but not on the correlation functions themselves.
This observation allows us to extend our analysis also
for the case of perturbative QFT on pseudo-Riemann manifolds.
\end{Remark}

\medskip

One can additionally prove by induction in $n$ that
\beq\label{eq3.14ne1}
\Jj_m \circ \Jj_{n-m} \, = \, \Jj_{n-m} \circ \Jj_m
\eeq
for $m<n$, where the composition $\Jj_m \circ \Jj_{n-m}$
is given by Eq.~(\ref{eq3.3ne2}) and similarly,
$\Jj_{n-m} \circ \Jj_m$ is the composition
\beqa\label{eq3.3ne22}
&& \DP_t \Bigl(F_m \bigl(\R^D\bigr) \times F_{n-m}\bigl(\R^D\bigr)
\times \mathop{\prod}\limits_r F_{n_r} \bigl(\R^{D_r}\bigr)\Bigr)
\nn
&& \hspace{12pt} \mathop{\longrightarrow}\limits^{\Jj_m} \,\,
\DP_t \Bigl(\R^{Dm} \times \R^{D(n-m)}
\times \mathop{\prod}\limits_r F_{n_r} \bigl(\R^{D_r}\bigr)\Bigr) \,
\nn
&& \hspace{10pt} \mathop{\longrightarrow}\limits^{\Jj_{n-m}} \,
\DP_t \Bigl(F_m \bigl(\R^D\bigr)
\times \mathop{\prod}\limits_r F_{n_r} \bigl(\R^{D_r}\bigr)\Bigr) \,.
\eeqa
The meaning of (\ref{eq3.14ne1}) is that subsequent extensions of a distribution
having singularities with respect to different groups of variables commute.
The proof of (\ref{eq3.14ne1}) can be done by induction in $n$
and using Eq.~(\ref{eq3.8ne2}).

Due to Eq.~(\ref{eq3.14ne1}) one can rewrite (\ref{eq3.8ne2}) in a more symmetric form:
\beq\label{eq3.16ne1}
\Jp_n\bigl(u\bigr)
\, := \,
\frac{1}{2} \, \mathop{\sum}\limits_{\mathop{}\limits^{S \, \subset \, \Nn}_{S \, \neq \, \emptyset}}
\Jj_S \circ \Jj_{S^c}
\bigl(
\chi_{\{S,S^c\}} \hspace{1pt} u
\bigr)
\vrestr{12pt}{\bigl(\R^{Dn}\bigr\backslash\Di_n\bigr) \times \mathop{\prod}\limits_r F_{n_r} \bigl(\R^{D_r}\bigr)} \,.
\eeq

Another corollary for $\Jj_n$ is that they satisfy the relations
\beqa\label{eq3.17ne1}
\Jj_n \Bigl(\di_{z_r^\xi} \hspace{1pt} u(\xx,\zz_1,\dots,\zz_s)\Bigr)
\podr = \,
\di_{z_r^\xi} \hspace{1pt} \Jj_n \bigl(u(\xx,\zz_1,\dots,\zz_s)\bigr) \,,
\\ \label{eq3.18ne1}
\Jj_n
\left(\raisebox{14pt}{\hspace{-2pt}}\right.
\mathop{\sum}\limits_{k \, = \, 1}^n \di_{x^{\mu}_k} \hspace{1pt} u(\xx,\zz_1,\dots,\zz_s)
\left.\raisebox{14pt}{\hspace{-2pt}}\right)
\podr = \,
\mathop{\sum}\limits_{k \, = \, 1}^n \di_{x^{\mu}_k} \hspace{1pt}
\Jj_n \bigl(u(\xx,\zz_1,\dots,\zz_s)\bigr) \,
\qquad
\eeqa
for $u \in \DP_t \Bigl(F_n \bigl(\R^D\bigr) \times \mathop{\prod}\limits_r F_{n_r} \bigl(\R^{D_r}\bigr)\Bigr)$,
$\eta=1,\dots,N_r$, $r=1,\dots,s$ and $\mu = 1,\dots,D$.
This is proven by using the properties (\ref{eq2.11ne31}) of $\II_N$.

In what follows we call the maps $\Jj_n$ \textbf{renormalization maps}.
For the sake of simplicity we shall consider them only as linear maps of the form
\beq\label{eq3.19ne2}
\Jj_n :
\DP_t \bigl(F_n \bigl(\R^D\bigr)\bigr)
\to
\DP_t \bigl(\R^{Dn}\bigr)
\,,
\eeq
and when we apply $\Jj_n$ on spaces of distributions depending on additional variables,
for instance,
on $\DP_t \Bigl(F_n \bigl(\R^D\bigr) \hspace{0pt} \times \mathop{\prod}\limits_r F_{n_r} \bigl(\R^{D_r}\bigr)\Bigr)$,
we shall consider these maps as
$\DP_t \Bigl(\mathop{\prod}\limits_r F_{n_r} \bigl(\R^{D_r}\bigr)\Bigr)$--linear.

Furthermore, due to Eq.~(\ref{eq3.18ne1})
we can consider only renormalization of
distributions depending on the relative distances
\beq\label{eq3.20ne2}
\x_{jk} \, = \, \x_j - \x_k \,.
\eeq
This is related to the projection
\beq\label{eq3.21ne31}
F_n \bigl(\R^D\bigr) \to F_{n-1} \bigl(\R^D \mz\bigr)
: \bigl(\x_1,\dots,\x_n\bigr) \mapsto \bigl(\x_{1n},\dots,\x_{n-1,n}\bigr)
\eeq
under which the total diagonal $\Di_n$ is mapped on $\{0\}$ and $\DI_n$ is projected on the cone
\beq\label{DIRED}
\DIR_{n-1} \, := \, \DI_{n-1} \cup \mathop{\bigcup}\limits_{k \, = \, 1}^{n-1} \{\y_k \in \R^D : \y_k \, = \, 0\} \,.
\eeq
We denote
\beq\label{eq3.22ne31}
\JJJ_n \bigl(v\bigr) \, := \, \Jj_n \bigl(u\bigr)
\eeq
for $u = v \bigl(\x_{1n},\dots,\x_{n-1,n}\bigr) \in \DP_t \bigl(F_{n-1} \bigl(\R^D \mz\bigr)\bigr)$,
and thus we obtain linear maps:
\beq\label{eq3.23ne31}
\JJJ_n : \CI_t \Bigl(F_{n-1} \bigl(\R^D \mz\bigr)\Bigr) \to \DP \bigl(\R^{D(n-1)}\bigr) \,
\eeq
restricting additionally our analysis to the vector space
\beqa\label{eq5.1ne41}
\hspace{-10pt}
\CI_t \Bigl(F_{n-1}\bigl(\R^D \mz\bigr)\Bigr) := \podr
\CI \Bigl(F_{n-1}\bigl(\R^D \mz\bigr)\Bigr) \cap
\DP_t \Bigl(F_{n-1}\bigl(\R^D \mz\bigr)\Bigr)
\nn
\hspace{-10pt}
= \podr\!
\mathop{\bigcup}\limits_{L \, \in \, \Z} \,
\CI_{(L)} \Bigl(F_{n-1}\bigl(\R^D \mz\bigr)\Bigr)
,
\\
\hspace{-10pt}
\CI_{(L)} \Bigl(F_{n-1}\bigl(\R^D \bigr\backslash\{0\}\bigr)\Bigr) := \podr
\CI \Bigl(F_{n-1}\bigl(\R^D \bigr\backslash\{0\}\bigr)\Bigr) \cap
\DP_{(L)} \Bigl(F_{n-1}\bigl(\R^D \bigr\backslash\{0\}\bigr)\Bigr)\,,
\nonumber
\eeqa
since this case is of main concern for us.
The maps $\JJJ_n$ satisfy analogous properties of those of $\Jj_n$ from Theorem~\ref{Th3.1ne1}
and the most important ones for us are
\beqa\label{eq3.24ne31}
&
\JJJ_n \bigl(u\bigr) \vrestr{12pt}{F_{n-1} \bigl(\R^D \mz\bigr)} \, = \, u \,,
& \\ \label{eq3.25ne31} &
\scdeg \, \JJJ_n \bigl(u\bigr) \, \leqslant \, \scdeg \, u \,,
& \\ \label{eq3.26ne31} &
\JJJ_n \bigl(f u\bigr) \, = \, f \, \JJJ_n \bigl(u\bigr)\,
&
\eeqa
for $u \in \CI_t \Bigl(F_{n-1} \bigl(\R^D \mz\bigr)\Bigr)$ and
$f \in \CI \bigl(\R^{D(n-1)}\bigr)$.
Finally, the maps $\JJJ_n$ have also an
inductive
construction in the form
\beq\label{eq3.27ne31}
\JJJ_n \, = \, \II_{D(n-1)} \circ \JPJ_n
\qquad (n>2), \qquad
\JJJ_2 \, = \, \II_D
 \,,
\eeq
where $\JPJ_n$ is determined by $\JJJ_2,\dots,\JJJ_{n-1}$.

\Section{Cohomology of renormalization maps}{se4ne1}
\setcntrs

As we have seen in the previous section,
the renormalization can be done in such a way that it commutes with the multiplication by smooth functions.
But on the other hand, it is not possible to make it commuting with the partial derivatives
(or, vector fields).
We study in this section how the renormalization ambiguity affects
the deviation from the latter commutativity.

Let us introduce the linear maps
\beqa\label{eq4.1ne1}
\hspace{-20pt}
\Lccl_{n;\,k,\,\mu} \podr := \, \di_{x^{\mu}_k} \circ \JJJ_n - \JJJ_n \circ \di_{x^{\mu}_k}
\, =: \, \bigl[\di_{x^{\mu}_k}, \JJJ_n\bigr] \,,
\nn
\hspace{-20pt}
\Lccl_{n;\,k,\,\mu} \podr :
\CI_t \Bigl(F_{n-1} \bigl(\R^D \mz\bigr)\Bigr)
\to
\DP_t \bigl[\DIR_{n-1}\bigr]
\,,
\eeqa
for $k=1,\dots,n-1$ and $\mu=1,\dots,D$
(recall that $x^{\mu}_k$ are now relative coordinates according to
Eqs.~(\ref{eq3.20ne2}) and (\ref{eq3.21ne31})).
By the representation
$\JJJ_n$ $=$ $\II_{D(n-1)} \circ \JPJ_n$
(\ref{eq3.27ne31}) we can represent $\Lccl_{n;\, k,\, \mu}$
in the form
\beqa\label{eq4.2ne1}
\Lccl_{n;\, k,\, \mu} \, = \podr
\Sccl_{n;\, k,\, \mu} \, + \, \widehat{\Lccl}_{n;\, k,\, \mu}
\qquad (n>2), \qquad \Lccl_{2;\, k,\, \mu} \, \equiv \, \Sccl_{2;\, k,\, \mu}
\, , \quad
\\ \label{eq4.3ne1}
\Sccl_{n;\, k,\, \mu} \, := \podr
\bigl[\di_{x_k^{\mu}},\II_{D(n-1)}\bigr] \circ \JPJ_n
\qquad (n>2)
\, , \quad
\\ \label{eq4.4ne1}
\widehat{\Lccl}_{n;\, k,\, \mu} \, := \podr
\II_{D(n-1)} \circ \bigl[\di_{x_k^{\mu}},\JPJ_n\bigr]
\qquad (n>2)
\, . \quad
\eeqa
Then the maps $\widehat{\Lccl}_{n;\, k,\, \mu}$ are determined by the renormalization
induction
and thus,
the only new information is contained in
\beq\label{eq4.5ne1}
\Sccl_{n;\, k,\, \mu} :
\CI_t \Bigl(F_{n-1} \bigl(\R^D \mz\bigr)\Bigr)
\to
\DP \bigl[0\bigr]
\eeq
$\DP [0]$ being the space of distributions supported at $0 \in \R^{D(n-1)}$.

The maps $\Sccl_{n;\, k,\, \mu}$ satisfy the ``differential equations'':
\beqa\label{eq4.6ne2}
\podr \!\!
\bigl[\di_{x_k^{\mu}},\Sccl_{2;\, j,\, \nu}\bigr]
-
\bigl[\di_{x_j^{\nu}},\Sccl_{2;\, k,\, \mu}\bigr] \, = \, 0 \,,
\nn
\podr \!\!
\bigl[\di_{x_k^{\mu}},\Sccl_{n;\, j,\, \nu}\bigr]
-
\bigl[\di_{x_j^{\nu}},\Sccl_{n;\, k,\, \mu}\bigr]
\\\ \podr \!\! \hspace{20pt}
= \,
- \, \bigl[\di_{x_k^{\mu}},\II_{D(n-1)}\bigr] \circ \bigl[\di_{x_j^{\nu}},\JPJ_n\bigr]
\, + \,
\bigl[\di_{x_j^{\nu}},\II_{D(n-1)}\bigr] \circ \bigl[\di_{x_k^{\mu}},\JPJ_n\bigr]
\, \quad (n > 2)
\nonumber
\eeqa
for $j,k=1,\dots,n-1$, $\mu,\nu=1,\dots,D$,
which are derived by a straightforward computation.
We shall characterize $\Sccl_{n;\, k,\, \mu}$ by these equations.
Before that, let us point out that \textit{the right hand side of (\ref{eq4.6ne2})
is determined by the renormalization induction}.
The reason is that the values of
$\bigl[\di_{x_j^{\nu}},\JPJ_n\bigr]$
are distributions supported at the rough diagonal $\DIR_{n-1}$ and then,
$\bigl[\di_{x_k^{\mu}},\II_{D(n-1)}\bigr]$
act on functions whose nontrivial dependence is in less than $n-1$ relative distances.
Thus, we consider (\ref{eq4.6ne2})
as equations for $\{\Sccl_{n;\, k,\, \mu}\}_{k,\, \mu}$ for fixed $n$,
whose right hand side is determined by $\{\Sccl_{m;\, k,\, \mu}\}_{k,\, \mu}$
for $m=2,\dots,n-1$.

Now let
\beq\label{eq4.7ne2}
\Sccl'_{n;\, k,\, \mu} \, := \, \Sccl_{n;\, k,\, \mu} +
\bigl[\di_{x_k^{\mu}},\RDF_n\bigr]
\eeq
($k=1,\dots,n-1$, $\mu=1,\dots,D$)
for some linear map
\beq\label{eq4.8ne2}
\RDF_n :
\CI_t \Bigl(F_{n-1} \bigl(\R^D \mz\bigr)\Bigr)
\to
\DP \bigl[0\bigr] \,.
\eeq
Then $\Sccl'_{n;\, k,\, \mu}$ also satisfy Eqs.~(\ref{eq4.6ne2})
but we shall see now that they correspond to another renormalization map
$\JJJ_n'$ $:$ $\CI_t \Bigl(F_{n-1} \bigl(\R^D \mz\bigr)\Bigr) \to \DP \bigl(\R^{D(n-1)}\bigr)$.
Since $\JPJ_n$ is an injection
$\CI_t \Bigl(F_{n-1} \bigl(\R^D \mz\bigr)\Bigr)$
$\hookrightarrow$
$\DP_t \bigl(\R^{D(n-1)} \mz\bigr)$
then there exists a linear map
\beq\label{eq4.9ne2}
\RDF'_n :
\DP_t \bigl(\R^{D(n-1)} \mz\bigr)
\to
\DP \bigl[0\bigr]
\eeq
such that
\beq\label{eq4.10ne2}
\RDF_n \, = \, \RDF_n' \circ \JPJ_n \,.
\eeq
Moreover, since
$\DP \bigl[\DIR_{n-1}\bigr] \cap \JPJ_n \Bigl(\CI_t \Bigl(F_{n-1} \bigl(\R^D \mz\bigr)\Bigr)\Bigr)$
$=$ $\{0\}$ we can additionally choose $\RDF'_n$ in such a way that
\beq\label{eq-qqq}
\RDF'_n \vrestr{12pt}{\DP \bigl[\DIR_{n-1}\bigr]} \, = \, 0 \,.
\eeq
Consider a new primary renormalization map
\beq\label{eq4.11ne2}
\II'_{D(n-1)} \, = \, \II_{D(n-1)} \, + \, \RDF_n' \,.
\eeq
Due of Eq.~(\ref{eq-qqq}) we have for it
\beq\label{eq4.12ne2}
\bigl[\di_{x_k^{\mu}},\II'_{D(n-1)}\bigr] \circ \JPJ_n
\, = \,
\Sccl'_{n;\, k,\, \mu} \, - \,
\RDF_n' \circ \bigl[\di_{x_k^{\mu}},\JPJ_n\bigr]
\, = \, \Sccl'_{n;\, k,\, \mu}
\,,
\eeq
since the image of $\bigl[\di_{x_k^{\mu}},\JPJ_n\bigr]$ is contained in the space $\DP \bigl[\DIR_{n-1}\bigr]$.
We obtained that $\Sccl'_{n;\, k,\, \mu}$ correspond by Eq.~(\ref{eq4.3ne1}) to another renormalization map
$\JJJ_n'$ $=$ $\II_{D(n-1)}' \circ \JPJ_n$.

Thus, we have seen that the maps $\Sccl_{n;\, k,\, \mu}$, for fixed $n$,
are characterized by
solutions of Eqs.~(\ref{eq4.6ne2}) modulo ``exact $1$--cocycles''.
But in general, the solution of (\ref{eq4.6ne2}) is unique up to a closed $1$--cocycle,
i.e., maps $c_{n;\, k,\, \mu}$ that satisfy the equations:
\beq\label{eq4.14ne2}
\bigl[\di_{x_j^{\nu}},c_{n;\, k,\, \mu}\bigr]
-
\bigl[\di_{x_k^{\mu}},c_{n;\, j,\, \nu}\bigr] \, = \, 0 \,
\eeq
(for $k=1,\dots,n-1$, $\mu=1,\dots,D$).
We shall see in Sect.~\ref{se5ne1} that the related cohomology is always finite dimensional:
it is equal to the de Rham cohomology $H^{D(n-1)-1} \Bigl(F_{n-1}\bigl(\R^D \mz\bigr)\Bigr)$.
In particular, it is zero for $n > 2$, since then
$H^{D(n-1)-1} \Bigl(F_{n-1}\bigl(\R^D\mz\bigr)\Bigr)$ $=$ $\{0\}$ (Theorem~\ref{Th5.3ne31}).
In this way we obtain the following important result.

\medskip

\begin{Theorem}\label{Th4.1qq1}
Let $\Sccl_{2;\, 1,\, \mu}$ for $\mu = 1,\dots,D$
be fixed by (\ref{eq4.3ne1}) for some renormalization map $\JJJ_2$.
Then every solution $\{\Sccl_{n;\, k,\, \mu}\}_{k,\, \mu}$ of Eqs.~(\ref{eq4.6ne2}) for $n>2$
corresponds by Eqs.~(\ref{eq4.2ne1})--(\ref{eq4.4ne1})
to some renormalization maps $\JJJ_3$, $\dots$, $\JJJ_n$.
\end{Theorem}

\medskip

We proceed in this section with a technical simplification for computing $\Sccl_{n;k,m}$.
Note that all these maps, together with $\RDF_n$, are maps of the form
\beq\label{eq4.15ne3}
\phi :
\CI_t \Bigl(F_{n-1} \bigl(\R^D \mz\bigr)\Bigr)
\to
\DP \bigl[0\bigr]
\eeq
and they satisfy the commutation relations
\beq\label{eq4.16ne2}
x_k^{\mu} \circ \phi - \phi \circ x_k^{\mu} \, = \, 0
\eeq
for all $k=1,\dots,n-1$ and $\mu=1,\dots,D$.
Furthermore, all the ``commutators'':
\beq\label{eq4.17ne2}
\bigl[\di_{x_k^{\mu}}, \phi\bigr] \, := \,
\di_{x_k^{\mu}} \circ \phi - \phi \circ \di_{x_k^{\mu}}
\eeq
are also maps of this class.
We shall see now that maps of such a type~(\ref{eq4.15ne3}), (\ref{eq4.16ne2})
can be determined just by linear functionals on
$\CI_t \Bigl(F_{n-1} \bigl(\R^D \mz\bigr)\Bigr)$.
We consider this construction in a more general situation.

\medskip

\begin{Definition}\label{Df3.1zz11}
Let us introduce the associative algebra $\DO{N}$ of all linear partial differential operators over $\R^N$
with polynomial coefficients\footnote{%
or instead of that, we can consider everywhere smooth coefficients, but for us polynomials will be enough.}.
A $\Z$--\textbf{filtered} $\DO{N}$--module is a module $\SCi$ of $\DO{N}$,
which is endowed with an increasing filtration
\beq\label{eq3.9ii1}
\SCi \, = \, \mathop{\bigcup}\limits_{L \, \in \, \Z} \SCi_{(L)}
\,,\quad
\SCi_{(L)} \, \subseteq \, \SCi_{(L+1)} \,,
\eeq
such that for every $A \in \DO{N}$ and $u \in \SCi$ we have
\beq\label{eq3.10pp1}
\scdeg \, A u \, \leqslant \, \scdeg \, A \, + \, \scdeg \, u \,,
\eeq
where
\beq\label{missing}
\scdeg \, u \, := \, \min \, \bigl\{L : u \in \SCi_{(L)}\bigr\}
\eeq
and the scaling degree of a differential operator is defined by:
\beq\label{eq4.18ne3}
\scdeg
\left(\raisebox{18pt}{\hspace{-2pt}}\right.
\mathop{\sum}\limits_{\rr \, \in \, \N_0^N}
f_{\rr} (\x) \, \di^{\rr}
\left.\raisebox{18pt}{\hspace{-2pt}}\right)
\, = \,
\mathop{\max}\limits_{\rr \, \in \, \N_0^N}
\, \bigl\{ |\rr| + \scdeg \, f_{\rr} \bigr\} \,.
\eeq
\end{Definition}%

\medskip

Clearly, $\CI_t \Bigl(F_{n-1} \bigl(\R^D \mz\bigr)\Bigr)$ is a $\Z$--filtered $\DO{D(n-1)}$--module.

\medskip

\begin{Proposition}\label{Th4.2xx2}
Let $\SCi$ be a $\Z$--filtered $\DO{N}$--module and let
$\phi : \SCi \to \DP \bigl[0\bigr]$ $($for $0 \in \R^N)$
satisfy $\bigr[ x^{\xxi} , \phi \bigl] = 0$.
Expand $\phi$ in delta functions and their derivatives:
\beq\label{eq4.12xx5}
\phi \, = \, \mathop{\sum}\limits_{\rr \, \in \, \N_0^N} \ \frac{1}{\rr!} \
\delta^{(\rr)} (\xx)
\, \Phi_{\rr}
\, , \quad
\Phi_{\rr} : \SCi \to \R \,.
\eeq
Then the assignment
\beq\label{eq4.23ne41}
\phi \, \mapsto \, \Phi_0
\eeq
is injective and
$\phi$ is determined by $\Phi_0$ by the formula
\beq\label{eq4.15new}
\Phi_{\rr} \, = \, (-1)^{|\rr|} \, \Phi_0 \circ \xx^{\rr}
\,.
\eeq%
Furthermore, under this assignment
\beq\label{eq4.24ne2}
\text{if} \quad \
\phi \mapsto \Phi_0
\quad \ \text{then} \quad \
\bigl[\di_{x^{\xi}},\phi\bigr] \mapsto - \Phi_0 \circ \di_{x^{\xi}} \,.
\eeq
\end{Proposition}%

\medskip

\begin{Proof}
The condition
$\bigr[ x^{\xxi} , \phi \bigl] = 0$
implies the recursive relation
$-(r_{\xi}+1)$ $\Phi_{\rr+\ee_{\xxi}}$ $=$ $\Phi_{\rr} \circ x^{\xxi}$,
from which we get (\ref{eq4.15new}).
Hence, $\phi \mapsto \Phi_0$ is injective.
Equation~(\ref{eq4.24ne2}) is obtained similarly.
\end{Proof}

\medskip

\begin{Definition}\label{Df4.3xx4}
For a $\DO{N}$--module $\SCi$ the \textbf{dual} module is the algebraic dual space $\DSCi$ endowed with
the conjugate action of $\DO{N}$: for $\Phi \in \DSCi$ and $\xxi=1,\dots,N$ the conjugate actions
of $x^{\xxi}$ and $\di_{x^{\xxi}}$ are
$x^{\xxi} \bigl(\Phi\bigr)$ $:=$ $\Phi \circ x^{\xxi}$ and
$\di_{x^{\xxi}} \bigl(\Phi\bigr)$ $:=$ $-\Phi \circ \di_{x^{\xxi}}$,
respectively.
\end{Definition}

\medskip

Thus, under the above definition Eq.~(\ref{eq4.24ne2}) reads:
\beq\label{eq4.25ne6}
\text{if} \quad \
\phi \mapsto \Phi_0
\quad \ \text{then} \quad \
\bigl[\di_{x^{\xi}},\phi\bigr] \mapsto \di_{x^{\xi}} \, \Phi_0 \,.
\eeq

Let us also denote
\beqa\label{eq4.26ne31}
\hspace{-5pt}
\RM \bigl(\SCi\bigr) \, :=
\Bigl\{ \podr\!\!
\phi : \,
\phi \text{ linearly maps } \SCi \text{ to } \DP \bigl[0\bigr],\
\bigl[x^{\xi},\phi\bigr] = 0 \ (\forall \xi = 1,\dots,N),\,
\nn \podr\!\!
\exists \, L \in \Z \text{ such that } \scdeg \, \phi (u) \leqslant L + \scdeg \, u \ (\forall u \in \SCi)
\Bigr\} \,.
\eeqa

\medskip

\begin{Proposition}\label{Pr4.3xx1}
The image of $\RM \bigl(\SCi\bigr)$ under the map~(\ref{eq4.23ne41}) is the vector space:
\beq\label{eq4.16xx6}
\SCi^{\du}
\, := \,
\mathop{\bigcup}\limits_{M \, \in \, \Z} \, \SCi_{(M)}^{\perp}
\ \ ( \, \subseteq \DSCi)
\, , \quad
\SCi_{(M)}^{\perp} \, := \,
\Bigl\{
\Phi \in \DSCi
:
\Phi\vrestr{10pt}{\SCi_{(M)}} = 0
\Bigr\}
\,.
\eeq%
The space $\SCi^{\du}$ is invariant under the action of
$\di_{x^{\xxi}}$ on $\DSCi$ ($\xxi=1,\dots,N$).
\end{Proposition}%

\medskip

\begin{Proof}
First, let $\phi \in \RM \bigl(\SCi\bigr)$ and let
$\scdeg \, \phi (u) \leqslant L + \scdeg \, u$.
Then $\Phi_0 \in \SCi_{(M)}^{\perp}$ if $M+L < N$.

Conversely, let
$\Phi \in \SCi^{\du}$ and
define by (\ref{eq4.12xx5}) and (\ref{eq4.15new}) with $\Phi_0 = \Phi$ a linear map
$\phi : \SCi \to \DP [0]$.
Note that the sum in (\ref{eq4.12xx5}) is always finite when we apply it on an element of $\SCi$,
since $\Phi \in \SCi_{(L)}^{\perp}$ for some $L \in \Z$ and $\xx^{\rr} \SCi_{(M)} \subseteq \SCi_{M-|\rr|}$
for every $M \in \Z$.
The latter also implies that if $\scdeg \, u = M$ then $\scdeg \, \phi \bigl(u\bigr)$ $\leqslant$ $N+K$,
where $L = M-K$.
Hence, $\phi \in \RM \bigl(\SCi\bigr)$ since the equations $\bigl[x^{\xxi},\phi\bigr]=0$ ($\xxi=1,\dots,N$)
follow as in the proof of Proposition~\ref{Th4.2xx2}.
By the construction $\phi$ is mapped on $\Phi$
via the assignment (\ref{eq4.15new}).
\end{Proof}

\medskip

\begin{Remark}\label{Rm4.2bb1}
The $\DO{N}$--module $\SCi^{\du}$ becomes also endowed with an increasing
$\Z$--filtration if we set $\SCi^{\du}_{(M)}$ $:=$ $\SCi_{(-M)}^{\perp}$.
\end{Remark}

\medskip

\begin{Definition}\label{Dfx1}
The above propositions suggest to introduce the analog of the \textit{de Rham complex}
for an arbitrary $\DO{N}$--module $\SCi$.
It is the
complex:
\beq\label{eq4.20ww6}
\{0\}
\mathop{\longrightarrow}\limits^{d} \,
\OM{0} \bigl(\SCi\bigr)
\, \mathop{\longrightarrow}\limits^{d} \,
\OM{1} \bigl(\SCi\bigr)
\, \mathop{\longrightarrow}\limits^{d} \,\, \cdots \,\, \mathop{\to}\limits^{d} \,
\OM{N} \bigl(\SCi\bigr)
\, \mathop{\longrightarrow}\limits^{d} \, \{0\}\,,
\eeq
where
\beq\label{eq4.20ww5}
\OM{0} \bigl(\SCi\bigr) \, \equiv \, \SCi
\, , \quad
\OM{m} \bigl(\SCi\bigr)
\, := \,
\Lambda^m \bigl(\R^N\bigr) \otimes
\SCi \,,
\eeq
and $\Lambda^m \bigl(\R^N\bigr)$ stands for the $m$th antisymmetric power of $\R^N$.
Thus, the elements of $\OM{m} \bigl(\SCi\bigr)$
are represented by sequences $\THETA$ $=$ $\bigl(\Theta_{\xxi_1,\dots,\xxi_m}\bigr)$ with
coefficients
$\Theta_{\xxi_1,\dots,\xxi_m}$ $\in$
$\SCi$
for $\xxi_1,\dots,\xxi_m = 1,\dots,N$, which are antisymmetric,
$\Theta_{\xxi_1,\dots,\xxi_m}$ $=$
$(-1)^{\text{sgn} \, \sigma}$ $\Theta_{\xxi_{\sigma_1},\dots,\xxi_{\sigma_m}}$.
The differential of $\Theta_{\xxi_1,\dots,\xxi_m}$ is
\beq\label{eq4.21ww5}
\bigl(d \THETA\bigr)_{\xxi_1,\dots,\xxi_{m+1}}
\, = \,
\mathop{\sum}\limits_{\ell \, = \, 1}^m \,
(-1)^{\ell+1} \,
\di_{x^{\xxi_{\ell}}} \,
\Theta_{\xxi_1,\dots,\widehat{\xxi_{\ell}},\dots,\xxi_{m+1}} \,.
\eeq
Denote by $\HOM{m} \bigl(\SCi\bigr)$, for $m=0,\dots,N$, the cohomology group of the complex (\ref{eq4.20ww6}):
\beqa\label{eq4.22ww5}
&& \hspace{50pt}
\HOM{m} \bigl(\SCi\bigr) \, := \,
\CLO{m} \bigl(\SCi\bigr)
{\hspace{1pt}}\Bigl/{\hspace{1pt}}
\EXA{m} \bigl(\SCi\bigr)
\, , \quad
\nn &&
\CLO{m} \bigl(\SCi\bigr) \, := \,
\Ker \ d \bigl|_{\OM{m} \bigl(\SCi\bigr)}
\, , \quad
\EXA{m} \bigl(\SCi\bigr) \, := \,
d \bigl(\OM{m-1} \bigl(\SCi\bigr)\bigr) \,.
\qquad
\eeqa
We also set
\beqa\label{xxt}
\OM{m} \bigl(\SCi\bigr){}_{(M)} \, := \podr
\Lambda^m \bigl(\R^N\bigr) \otimes \SCi_{(M)}
\, , \quad
\nn
\CLO{m} \bigl(\SCi\bigr){}_{(M)} \, := \podr
\CLO{m} \bigl(\SCi\bigr) \cap
\OM{m} \bigl(\SCi\bigr){}_{(M)}
\, , \quad
\nn
\EXA{m} \bigl(\SCi\bigr){}_{(M)} \, := \podr
\EXA{m} \bigl(\SCi\bigr) \cap
\OM{m} \bigl(\SCi\bigr){}_{(M)} \,.
\eeqa%
\end{Definition}%

\Section{Reduction to de Rham cohomologies}{se5ne1}
\setcntrs

In Theorem \ref{Th4.1qq1} of the previous section
we have found cohomological equations (\ref{eq4.6ne2})
that characterize
the maps
$\Sccl_{n;\, k,\, \mu}$~(\ref{eq4.3ne1}), (\ref{eq4.5ne1})
up to a change of renormalization.
For the proof of this theorem
it was important to know what is the cohomology space
related to the left hand side of Eqs.~(\ref{eq4.6ne2})
and we claimed (without a proof) that it is zero for $n>2$.
By Propositions~\ref{Th4.2xx2} and \ref{Pr4.3xx1}
it follows that this cohomology space is exactly equal to
\beq\label{eq5.1qq1}
H^1 \Bigl(\CI_t \Bigl(F_{n-1}\bigl(\R^D \mz\bigr)\Bigr)\raisebox{10pt}{\hspace{-1pt}}^{\du}\hspace{1pt}\Bigr) \,.
\eeq
In this section we show that the space~(\ref{eq5.1qq1}) is isomorphic to the dual of
the de Rham cohomology space
\beq\label{eq5.2qq1}
H^{D(n-1)-1} \bigl(F_{n-1}\bigl(\R^D \mz\bigr)\bigr) \,.
\eeq
Then, from the known results about de Rham cohomologies of configuration spaces
(see Theorem~\ref{Th5.3ne31} below) it follows that (\ref{eq5.1qq1}) is zero for $n>2$.

First we start with an analog of the Poincar\'e lemma in the case of
de Rham cohomologies of $\DO{N}$--modules.
It uses partial inevitability of the Euler operator (vector field) on $\R^N$,
\beq\label{eq5.1qq11}
\xx \spr \di_{\xx} \, = \, \mathop{\sum}\limits_{\xxi \, = \, 1}^N \, x^{\xxi} \, \di_{x^{\xxi}} \,.
\eeq

\medskip

\begin{Proposition}\label{Pr5.1qq1}
Let $\SCi$ be a filtered $\DO{N}$--module, which is such that for some $M \in \Z$
the operator $\ell+\xx \spr \di_{\xx}$
it is invertible on $\SCi_{(M)}$ for every $\ell \in \N_0$.
Then every closed form with coefficients in $\SCi_{(M)}$ is exact, i.e.,
if $\THETA$ $\in$
$\OM{m} \bigl(\SCi\bigr){}_{(M)}$
then $d \hspace{1pt} \THETA$ $=$ $0$
implies that $\THETA = d \BBB$ for some $\BBB$ $\in$ $\OM{m-1} \bigl(\SCi\bigr)$.
\end{Proposition}

\medskip

\begin{Proof}
The proposition can be proven by using the $\DO{N}$--analog of the Po\-i\-n\-c\-a\-r\'e integration operator
\beq\label{eq5.2qq11}
\bigl(K \hspace{1pt} \THETA\bigr)_{\xxi_1,\dots,\xxi_{m-1}} \, = \,
\bigl(m+\xx \spr \di_{\xx}\bigr)^{-1} \,
\mathop{\sum}\limits_{\xxi \, = \, 1}^N \, x^{\xxi} \, \Theta_{\xxi,\xxi_1,\dots,\xxi_{m-1}}
\eeq
where $\THETA = \bigl(\Theta_{\xxi_1,\dots,\xxi_m}\bigr)$ $\in$
$\OM{m} \bigl(\SCi\bigr){}_{(M)}$ $=$
$\Lambda^m \bigl(\R^N\bigr) \otimes \SCi_{(M)}$.
Since the operators $m+\xx \spr \di_{\xx}$ commute with the derivatives $\di_{x^{\xxi}}$
then the operators $\bigl(m+\xx \spr \di_{\xx}\bigr)^{-1}$ also commute with $\di_{x^{\xxi}}$
on $\Lambda^m \bigl(\R^N\bigr) \otimes \SCi_{(M-1)}$.
Hence, we derive the identity $Kd + dK = \id$ from which the proposition follows.
\end{Proof}

\medskip

There is a natural pairing
\beq\label{eq4.27hh1}
\OM{m} \bigl(\SCi^{\du}\bigr)
\otimes
\OM{N-m} \bigl(\SCi\bigr)
\, \to \, \R
\, : \, \bigl(\OMEGA,\, \aLPHA \bigr) \, \mapsto \, \OMEGA\wed \bigl[\aLPHA\bigr]
\,,
\eeq
where $\OMEGA\wed \bigl[\aLPHA\bigr]$ means the action of $\OMEGA$ as a linear functional
under the external product $\wedge$.
Precisely, for $\OMEGA$ $=$ $a \otimes \Phi$ $\in$ $\Lambda^m \bigl(\R^N\bigr)$ $\otimes$ $\SCi^{\du}$
and $\aLPHA$ $=$ $b \otimes f$ $\in$ $\Lambda^{N-m} \bigl(\R^N\bigr)$ $\otimes$ $\SCi$
we set:
\beq\label{eq4.28hh1}
\OMEGA\wed \bigl[\aLPHA\bigr] \, := \,
\bigl(a \wedge b \bigr) \, \Phi \bigl[f\bigr] \,,
\eeq
where $a \wedge b$ is considered as an element of $\R$ $\cong$ $\Lambda^N \bigl(\R^N\bigr)$.
Then we have
\beq\label{eq4.29hh1}
\bigl(d \hspace{1pt} \OMEGA\bigr)\wed \bigl[\aLPHA\bigr] \, = \,
(-1)^{m+1} \, \OMEGA\wed \bigl[d \hspace{1pt} \aLPHA\bigr] \,
\eeq
for $\OMEGA$ $\in$ $\OM{m} \bigl(\SCi\bigr)$ and $\aLPHA$ $\in$ $\OM{N-m-1} \bigl(\SCi\bigr)$,
according to Eq.~(\ref{eq4.21ww5}).
We denote
\beq\label{eq5.6nmn1}
\OMEGA\wed \, : \, \OM{N-m} \bigl(\SCi\bigr) \, \to \, \R \, : \, \aLPHA \, \mapsto \, \OMEGA\wed \bigl[\aLPHA\bigr] \,.
\eeq
Note now that if $\OMEGA$ is closed then
$\OMEGA\wed \bigl[d\hspace{1pt}\aLPHA'\bigr] = 0$ for every $\aLPHA'$ $\in$ $\OM{N-m-1} \bigl(\SCi\bigr)$
and if $\OMEGA$ is exact then
$\OMEGA\wed \bigl[\aLPHA\bigr] = 0$ for every closed $\aLPHA$ $\in$ $\OM{N-m} \bigl(\SCi\bigr)$.
Hence, for closed $\OMEGA$ and $\aLPHA$ the number $\OMEGA\wed \bigl[\aLPHA\bigr]$
depends only on the cohomology classes of $\OMEGA$ and $\aLPHA$
and thus we obtain a natural linear map
\beq\label{eq4.30hh1}
\HOM{m} \bigl(\SCi^{\du}\bigr) \, \to \,
\bigl( \HOM{N-m} \bigl(\SCi\bigr) \bigr)^*
\,.
\eeq

\medskip

\begin{Proposition}\label{Pr5.2qq1}
Under the assumptions of Proposition~\ref{Pr5.1qq1} the above natural map (\ref{eq4.30hh1})
is an isomorphism $\HOM{m} \bigl(\SCi^{\du}\bigr)$ $\cong$ $\bigl(\HOM{N-m} \bigl(\SCi\bigr)\bigr)^*$.
\end{Proposition}

\medskip

\begin{Proof}
First we prove that (\ref{eq4.30hh1}) is surjective.
Let us have a linear functional
\beq\label{eq5.7jj6}
\CLO{N-m} \bigl(\SCi\bigr){\hspace{-1pt}}\Bigl/{\hspace{0pt}}\EXA{N-m} \bigl(\SCi\bigr)
\ ( = \HOM{N-m} \bigl(\SCi\bigr) )
\ \mathop{\longrightarrow}\limits^{\Omega'} \ \R \,.
\eeq
Then our task is to extend $\Omega'$ to a linear functional
\beq\label{eq5.8jj6}
\OM{N-m} \bigl(\SCi\bigr){\hspace{-1pt}}\Bigl/{\hspace{0pt}}\EXA{N-m} \bigl(\SCi\bigr)
\, \mathop{\longrightarrow}\limits^{\Omega''} \, \R
\eeq
such that $\Omega'' \circ \pi = \OMEGA\wed$ for an element $\OMEGA \in \OM{m} \bigl(\SCi^{\du}\bigr)$,
where $\pi$ is the natural projection
$\OM{N-m} \bigl(\SCi\bigr)$ $\to$
$\OM{N-m} \bigl(\SCi\bigr){\hspace{-1pt}}\Bigl/{\hspace{0pt}}\EXA{N-m} \bigl(\SCi\bigr)$.
It is always possible to extend $\Omega'$~(\ref{eq5.7jj6}) to some linear functional $\Omega''$~(\ref{eq5.8jj6})
and we point out also that every linear functional
$\Theta : \OM{N-m} \bigl(\SCi\bigr) \to \R$
is of a form $\THETA\wed$ for a unique $\THETA \in \OM{m} \bigl(\DSCi\bigr)$.
So, we have an element $\OMEGA \in \OM{m} \bigl(\DSCi\bigr)$ such that $\OMEGA\wed = \Omega'' \circ \pi$
and it remains only to achieve $\OMEGA \in \OM{m} \bigl(\SCi^{\du}\bigr)$.
The latter requires to impose further conditions on the extension $\Omega''$~(\ref{eq5.8jj6}).
We require that $\Omega''$ is zero on
$$
\pi \bigl( \OM{N-m} \bigl(\SCi\bigr){}_{(M)}\bigr) \, \equiv \,
\OM{N-m} \bigl(\SCi\bigr){}_{(M)}{\hspace{-1pt}}\Bigl/{\hspace{0pt}}\EXA{N-m} \bigl(\SCi\bigr)
$$
for some $M \in \Z$.
This is always possible since
\beqa
\podr
\Bigl(\OM{N-m} \bigl(\SCi\bigr){}_{(M)}{\hspace{-1pt}}\Bigl/{\hspace{0pt}}\EXA{N-m} \bigl(\SCi\bigr)\Bigr)
\cap
\Bigl(\CLO{N-m} \bigl(\SCi\bigr){\hspace{-1pt}}\Bigl/{\hspace{0pt}}\EXA{N-m} \bigl(\SCi\bigr)\Bigr)
\nn \podr \quad = \,
\CLO{N-m} \bigl(\SCi\bigr){}_{(M)}{\hspace{-1pt}}\Bigl/{\hspace{0pt}}\EXA{N-m} \bigl(\SCi\bigr)
\, = \, \{0\}
\nonumber
\eeqa
for $M$ chosen according to the assumptions of Proposition~\ref{Pr5.1qq1}.
In this way we have
$\OMEGA\wed\bigl[\aLPHA\bigr]=0$ if $\aLPHA \in \OM{N-m} \bigl(\SCi\bigr){}_{(M)}$
and hence, $\OMEGA \in \OM{m} \bigl(\SCi^{\du}\bigr)$.
Thus, the map (\ref{eq4.30hh1}) is surjective.

To prove that the map (\ref{eq4.30hh1}) is injective
assume that $\OMEGA \in \OM{m} \bigl(\SCi^{\du}\bigr)$ is such that
$\OMEGA\wed \bigl[\aLPHA\bigr] = 0$ for all $\aLPHA \in \CLO{N-m} \bigl(\SCi\bigr)$.
We should prove that $\OMEGA = d \THETA$ for $\THETA \in \OM{m-1} \bigl(\SCi^{\du}\bigr)$.
To this end we note first that
$\OMEGA\wed\vrestr{10pt}{\OM{N-m} \bigl(\SCi\bigr){}_{(M')}}$ $=$ $0$ for some $M' \in \Z$
and we set $M_0$ $=$ $\min\{M,M'\}$, where $M \in \Z$
is the integer from the assumptions of Proposition~\ref{Pr5.1qq1}.
Thus, $\OMEGA\wed\vrestr{10pt}{\OM{N-m} \bigl(\SCi\bigr){}_{(M_0)}}$ $=$ $0$.
Now consider the short exact sequence
\beqa\label{eq5.9jj6}
&&
0 \, \to \, \CLO{N-m} \bigl(\SCi\bigr){\hspace{-1pt}}\Bigl/{\hspace{0pt}}\CLO{N-m} \bigl(\SCi\bigr){}_{(M_0-1)}
\, \hookrightarrow \,
\OM{N-m} \bigl(\SCi\bigr){\hspace{-1pt}}\Bigl/{\hspace{0pt}}\OM{N-m} \bigl(\SCi\bigr){}_{(M_0-1)}
\nn && \hspace{50pt}
\mathop{\to}\limits^{d} \,
\EXA{N-m+1} \bigl(\SCi\bigr){\hspace{-1pt}}\Bigl/{\hspace{0pt}}\EXA{N-m+1} \bigl(\SCi\bigr)_{(M_0)} \, \to \, 0 \,
\,,
\eeqa
which is due to $\EXA{N-m+1} \bigl(\SCi\bigr)_{(M_0)}$
$=$ $d \, \OM{N-m} \bigl(\SCi\bigr)_{(M_0-1)}$
(Proposition~\ref{Pr5.1qq1}).
Then we obtain a linear functional
\beq\label{eq5.10jj1}
\Theta' : \EXA{N-m+1} \bigl(\SCi\bigr){\hspace{-1pt}}\Bigl/{\hspace{0pt}}\EXA{N-m+1} \bigl(\SCi\bigr)_{(M_0)} \to \R
\eeq
such that $\OMEGA\wed = \Theta' \circ \pi' \circ d$,
where $\pi'$ is the projection
$$
\EXA{N-m+1} \bigl(\SCi\bigr) \, \mathop{\longrightarrow}\limits^{\pi'} \,
\EXA{N-m+1} \bigl(\SCi\bigr){\hspace{-1pt}}\Bigl/{\hspace{0pt}} \EXA{N-m+1} \bigl(\SCi\bigr)_{(M_0)}\,.
$$
Finally, we extend $\Theta'$ to a linear functional $\Theta''$,
\beq\label{eq5.10jj8}
\Theta'' : \OM{N-m+1} \bigl(\SCi\bigr){\hspace{-1pt}}\Bigl/{\hspace{0pt}}\OM{N-m+1} \bigl(\SCi\bigr)_{(M_0)} \to \R
\eeq
(under the natural embedding
$$
\EXA{N-m+1} \bigl(\SCi\bigr){\hspace{-1pt}}\Bigl/{\hspace{0pt}}\EXA{N-m+1} \bigl(\SCi\bigr)_{(M_0)}
\, \hookrightarrow \,
\OM{N-m+1} \bigl(\SCi\bigr){\hspace{-1pt}}\Bigl/{\hspace{0pt}}\OM{N-m+1} \bigl(\SCi\bigr)_{(M_0)} \,)
$$
and setting $\THETA\wed := \Theta'' \circ \pi''$, where
$$
\OM{N-m+1} \bigl(\SCi\bigr) \, \mathop{\longrightarrow}\limits^{\pi''} \,
\OM{N-m+1} \bigl(\SCi\bigr){\hspace{-1pt}}\Bigl/{\hspace{0pt}} \OM{N-m+1} \bigl(\SCi\bigr)_{(M_0)}\,,
$$
we get by Eq.~(\ref{eq4.29hh1}) that
$\OMEGA$ $=$ $(-1)^{m+1} \, d \,\THETA$ and $\THETA \in \OM{m-1} \bigl(\SCi^{\du}\bigr)$
(i.e., $\OMEGA\wed\bigl[\aLPHA\bigr] = \THETA\wed\bigl[d\hspace{1pt}\aLPHA\bigr]$
for all $\aLPHA \in \OM{m-1} \bigl(\SCi\bigr)$).
Hence, $\THETA$ is exact in $\OM{m} \bigl(\SCi^{\du}\bigr)$ and thus, the map~(\ref{eq4.30hh1})
is also injective.
\end{Proof}

\medskip

\begin{Proposition}\label{Pr5.3jj1}
The $\DO{D(n-1)}$--modules
$\CI_t \Bigl(F_{n-1}\bigl(\R^D \mz\bigr)\Bigr)$
satisfy the assum\-p\-tions of Proposition~\ref{Pr5.1qq1} and hence,
we have natural isomorphism
\beq\label{eq.v2-1}
\HOM{m} \Bigl(\CI_t \Bigl(F_{n-1}\bigl(\R^D \mz\bigr)\Bigr)\raisebox{10pt}{\hspace{-1pt}}^{\du}\hspace{1pt}\Bigr)
\, \cong \,
\HOM{N-m} \Bigl(\CI_t \Bigl(F_{n-1}\bigl(\R^D \mz\bigr)\Bigr)\Bigr)^* \,.
\eeq
\end{Proposition}

\medskip

\begin{Proof}
Denote $\SCI_n$ $:=$ $\CI_t \Bigl(F_{n-1}\bigl(\R^D \mz\bigr)\Bigr)$.
We shall prove that
all the operators
$\ell+\xx \spr \di_{\xx}$, for $\ell=0,1,2,\dots$, are invertible
on the subspace of $\OM{m} \bigl(\SCI_n\bigr)$, which consists of
elements with negative scaling degree.
Indeed, if
$\THETA$ $=$ $\bigl(\Theta_{\xxi_1,\dots,\xxi_m} (\xx)\bigr)$ $\in$ $\OM{m} \bigl(\SCI_n\bigr)$
and $\scdeg \, \THETA < 0$
then for every $\xx \notin \DI_n$ the function
$\lambda^{-1}$
$\Theta_{\xxi_1,\dots,\xxi_m} (\lambda \hspace{1pt} \xx)$
is integrable for $\lambda \in \bigl(0,1\bigr)$.
Hence, we define $\bigl(\ell+\xx \spr \di_{\xx}\bigr)^{-1} \, \THETA$ by
\beq\label{eq5.12hh1}
\Bigl(\bigl(\ell+\xx \spr \di_{\xx}\bigr)^{-1} \, \THETA\Bigr)
\raisebox{-5pt}{\hspace{0pt}}_{\xxi_1,\dots,\xxi_m} \bigl(\xx\bigr)
\, = \,
\mathop{\int}\limits_{\hspace{-7pt}0}^{\hspace{5pt}1}
\lambda^{\ell-1} \,
\THETA_{\xxi_1,\dots,\xxi_m} \bigl(\lambda \hspace{1pt} \xx\bigr) \, d\hspace{1pt}\lambda
\,,
\eeq
where the right hand side defines an element of $\SCI_n$.
\end{Proof}

\medskip

Thus, we almost reduced the the cohomology spaces
$$
H^* \Bigl(\CI_t \Bigl(F_{n-1}\bigl(\R^D \mz\bigr)\Bigr)\raisebox{10pt}{\hspace{-1pt}}^{\du}\hspace{1pt}\Bigr)
$$
appearing in the analysis of renormalization ambiguity to de Rham cohomologies
$$
H^* \bigl(F_{n-1}\bigl(\R^D \mz\bigr)\bigr)
\, \equiv \,
H^* \Bigl(\CI \Bigl(F_{n-1}\bigl(\R^D \mz\bigr)\Bigr)\Bigr) \,.
$$
It only remains to point out that we have an equivalence
$$
H^* \Bigl(\CI_t \Bigl(F_{n-1}\bigl(\R^D \mz\bigr)\Bigr)\Bigr)
\, \cong \,
H^* \Bigl(\CI \Bigl(F_{n-1}\bigl(\R^D \mz\bigr)\Bigr)\Bigr) \,,
$$
i.e., if we work with forms with a tempered growth at the boundary $\DIR_{n-1}$
then this does not change the cohomology classes.
(The latter follows by the fact that every cohomology class in
$H^* \bigl(F_{n-1}\bigl(\R^D \mz\bigr)\bigr)$ has a representative belonging to
$\CI_t \Bigl(F_{n-1}\bigl(\R^D \mz\bigr)\Bigr)$, according to Theorem~\ref{Th5.3ne31} below.)

Now in order to explain the structure of $H^* \bigl(F_n\bigr)$ let us introduce the maps
\beqa\label{eq5.15ne31}
\hspace{-30pt}
\psi : \podr \Sr^{D-1} \,\to\, F_2 \bigl(\R^D\bigr) : \x \,\mapsto\, \bigl(\x,-\x\bigr) \,,
\\ \label{eq5.16ne31}
\hspace{-30pt}
\pi_{j_1,\dots,j_m}^n : \podr F_n \bigl(\R^D\bigr) \,\to\, F_m \bigl(\R^D\bigr)
: \bigl(\x_1,\dots,\x_n\bigr) \,\mapsto\, \bigl(\x_{j_1},\dots,\x_{j_m}\bigr) \,,
\eeqa
for $1 \leqslant j_1 < \cdots < j_m \leqslant n$.
Let $\alpha$ be a closed $(D-1)$--form such that $\psi^* \, \alpha$
is the volume form on $\Sr^{D-1}$ and set for $1 \leqslant j < k \leqslant n$:
\beq\label{eq5.17ne41}
\alpha_{j,k} \, := \, \bigl(\pi_{j,k}^{n}\bigr)^* \, \alpha \,,
\eeq
which is a closed $(D-1)$--form on $F_n \bigl(\R^D\bigr)$
and we denote by $[\alpha_{i,j}]$ the corresponding cohomology class
in $H^{D-1} \bigl(F_n \bigl(\R^D\bigr)\bigr)$.

\medskip

\begin{Theorem}\label{Th5.3ne31}
{\rm \cite{FH}}
Let $Q \, := \, \{0,\ee,\dots,q\,\ee\} \subset \R^D$ for $q=0,1,\dots$,
where $\ee$ is a nonzero vector in $\R^D$.
The spaces $H^r \bigl(F_n\bigl(\R^D \bigr\backslash Q\bigr)\bigr)$ are finite dimensional
and the only nonzero ones are for $r=s \, (D-1)$ with $s=1,\dots,n-1$.
The algebra $H^* \bigl(F_n\bigl(\R^D\bigr)\bigr)$ is a free algebra
with generators $[\alpha_{j,k}]$ for $1 \leqslant j < k \leqslant n$ and relations
\beqa\label{eq5.18ne41}
&
[\alpha_{j,k}]^2 \, = \, 0 \quad (j < k)
\,,
& \\ \label{eq5.19ne41} &
[\alpha_{j,\ell}] [\alpha_{k,\ell}] \, = \, [\alpha_{j,k}] [\alpha_{k,\ell}] - [\alpha_{j,k}] [\alpha_{j,\ell}]
\quad (j < k < \ell)
\,. &
\eeqa%
\end{Theorem}%

\medskip

We see that the maximal rank $r$ for which $H^r \bigl(F_{n-1}\bigl(\R^D \mz\bigr)\bigr) \neq \{0\}$ is $(n-1)(D-1)$.
Hence, $H^{D (n-1) -1} \bigl(F_{n-1}\bigl(\R^D \mz\bigr)\bigr)$ are always zero for~$n > 2$.

\Section{Cohomological equations. Outlook}{se6ne1}
\setcntrs

We conclude the paper with a discussion how the cohomological equations (\ref{eq4.6ne2})
can be further reduced to
$\CI_t \Bigl(F_{n-1} \bigr(\R^D \mz\bigl)\Bigr)\raisebox{10pt}{\hspace{-1pt}}^{\du}\hspace{1pt}$
according to the isomorphism provided by Propositions~\ref{Th4.2xx2} and \ref{Pr4.3xx1}.
By these results the maps
$\Sccl_{n;\, k,\, \mu}$ ($k=1,\dots,n-1$, $\mu =1,\dots,D$)
are characterized by linear functionals
$\sccl_{n;\, k,\, \mu}$ that belong to
$\CI_t \Bigl(F_{n-1} \bigr(\R^D \mz\bigl)\Bigr)\raisebox{10pt}{\hspace{-1pt}}^{\du}\hspace{1pt}$
(see Eqs.~(\ref{eq4.12xx5}) and (\ref{eq4.23ne41})).
Thus, we need to find the counterpart of the cohomological equations (\ref{eq4.6ne2}) for
the functionals $\sccl_{n;\, k,\, \mu}$.

To this end we first point out that $\sccl_{n;\, k,\, \mu}$ can be organized in one-forms
$\bsccl_n \in \OM{1} \Bigl(\CI_t \Bigl(F_{n-1} \bigr(\R^D \mz\bigl)\Bigr)\raisebox{10pt}{\hspace{-1pt}}^{\du}\hspace{1pt}\Bigr)$
(cf.~Definition~\ref{Dfx1})
and the cohomological equations for them are of the following general type:
\beqa\label{eq6xx}
d \hspace{1pt} \bsccl_2 \, = \podr 0,
\nn
d \hspace{1pt} \bsccl_n \, = \podr
\mathcal{F} \bigl[\bsccl_1,\dots,\bsccl_{n-1}\bigr]
\qquad (n>2) \,.
\eeqa
In order to derive the right hand side of (\ref{eq6xx}) in a simple explicit form
we further restrict the renormalization map $\JJJ_n$, for every $n=2,3,\dots$, to the subspace $\Ee_n$ of
$\CI_t \Bigl(F_{n-1} \bigr(\R^D \mz\bigl)\Bigr)$, which consists of all \textit{finite sums of products} of type
\beq\label{eq6.1n51}
u \, = \,
\left(\raisebox{16pt}{\hspace{-2pt}}\right.
\mathop{\prod}\limits_{1 \, \leqslant \, j \, < \, k \, \leqslant \, n-1} G_{j,k} (\x_j-\x_k)
\left.\raisebox{16pt}{\hspace{-5pt}}\right)
\left(\raisebox{16pt}{\hspace{-2pt}}\right.
\mathop{\prod}\limits_{\ell \, = \, 1}^{n-1} G_{\ell,n} (\x_{\ell})
\left.\raisebox{16pt}{\hspace{-5pt}}\right) \,
\eeq
where $G_{j,k},G_{\ell,n} \in \CI_t \bigr(\R^D \mz\bigl)$
(recall that $\x_k$ have now meaning of relative distances according to Eqs.~(\ref{eq3.20ne2}) and (\ref{eq3.21ne31})).
Thus, $\Ee_n$ is a $\Z$--filtered $\DO{D(n-1)}$--module (for $n=2,3,\dots$) and
we shall look for $\bsccl_n$
as an elements of the $\DO{D(n-1)}$--module $\OM{1}\bigl(\Ee_n^{\du}\bigr)$.
Let us introduce, for every proper subset $S$ $\subset$ $\Nn$ $(= \{1,\dots,n\})$,
maps $\Sccl_{S;\, k,\, \mu}$
($k=1,\dots,n-1$, $\mu=1,\dots,D$)
as acting on $u$~(\ref{eq6.1n51}) in the following way:
\beq\label{eq6.2ne61}
\Sccl_{S;\, k,\, \mu}
\bigl(u\bigr)
\, := \,
u_S^c
\, \cdot \,
\Sccl_{|S|;\, k,\, \mu}
\bigl(
u_S
\bigr)
\eeq
where
$u_S$ is the part of the two products in Eq.~(\ref{eq6.1n51}), which consists of all
$G_{j',k'}$ with $j'<k'$ and $j',k' \in S$, and
$u_S^c$ stands for the remaining part of the products.
The maps $\Sccl_{S;\, k,\, \mu}$ are considered as linear maps between the spaces:
\beq\label{eq6.2n51}
\Sccl_{S;\, k,\, \mu}
: \Ee_n \to \Ee_{\Nn / S} \otimes \DP [0] \,,
\eeq
where $\DP [0]$ is the space of distributions supported at $0 \in \R^{D(m-1)}$
(with respect to the variables related to $S \subset \Nn$),
and we have also set
\beq\label{eq6.1ne1.1}
\Nn \bigl/ S \, := \, S^c \cup \{\max \, S\} \,
\qquad
(\bigr|\Nn \bigl/ S\bigl| \, = \, n - |S| +1)\,,
\eeq
and
$\Ee_S$ (for any $S \subseteq \Nn$) stands for the space of finite sums of products
$\mathop{\prod}\limits_{\mathop{}\limits^{j,\, k \, \in \, S}_{j \, < \, k}} G_{j,k}$.
Under these conventions one can derive the equalities:
\beq\label{eq6.3n51}
\bigl[\di_{x_k^{\mu}}, \JPJ_n \bigr]
\, = \,
\mathop{\sum}\limits_{\mathop{}\limits^{S \, \subset \, \Nn}_{|S| \, \geqslant \, 2}}
\, \Bigl( \JPJ_{\Nn / S} \otimes \ID_{\DP [0]} \Bigr) \, \circ \, \Sccl_{S;\, k,\, \mu} \,.
\eeq
Equations~(\ref{eq6.3n51}) are derived under some additional conditions
on the renormalization maps $\JJJ_n$, which
break the permutation symmetry (cf., Theorem \ref{Th3.1ne1} (a))
but it can be always restored by a symmetrization at the end.

Let $\sccl_{S;\, k,\, \mu}$ correspond to $\Sccl_{S;\, k,\, \mu}$ under the assignment
(\ref{eq4.23ne41}) and the decomposition (\ref{eq4.12xx5}).
These maps $\sccl_{S;\, k,\, \mu}$ are now linear maps
\beq\label{eq6.7ne1}
\sccl_{S;\, k,\, \mu}
:
\Ee_n
\, \to \,
\Ee_{\Nn / S} \,
\eeq
and we collect them into a $1$--form $\bsccl_S$ similarly to $\bsccl_n$.
So, $\bsccl_S$ is basically $\bsccl_{|S|}$ but extended to
$\Ee_n$ as a collection of maps commuting with the multiplication
by functions that do not have singularities
with respect to $\x_j-\x_k$ for all $j,s \in S$.

Finally, under all the above considerations one can derive the following counterpart
of the cohomological equations (\ref{eq4.6ne2}):
\beqa\label{eq6.9ne1}
d \hspace{1pt} \bsccl_2 \, = \podr 0
\, , \qquad
\nn
d \hspace{1pt} \bsccl_n \, = \podr - \, \mathop{\sum}\limits_{\mathop{}\limits^{S \, \subset \, \Nn}_{|S| \, \geqslant \, 2}}
\bsccl_{\Nn / S} \wedge \bsccl_S
\qquad (n > 2)
\,.
\eeqa

We intend to study the cohomological equations (\ref{eq6.9ne1}) and their solutions in the future.

\medskip


\noindent
\textbf{Acknowledgments.}
I am grateful to Stefan Hollands
who triggered my interest to study various algebraic
structures related to the perturbative QFT.
I am also grateful to Ivan Todorov for a critical reading of the manuscript.
The results of this work
were presented at the 4th Vienna Central European Seminar on Particle Physics and Quantum Field Theory
(30 November -- 2 December, 2007).
The stay during the conference was supported by Austrian Federal Ministry of
Science and Research, the High Energy Physics Institute
of the Austrian Academy of Sciences and
the Erwin Schr\"odinger International Institute of Mathematical Physics.


\end{document}